\documentclass[11pt]{article}
\usepackage{graphicx}
\usepackage[table]{xcolor}
\usepackage{amsmath}
\usepackage{array,booktabs,siunitx}
\usepackage[utf8]{inputenc}
\usepackage{multirow}
\usepackage{fmtcount} 
\usepackage{array,multirow}
\newcolumntype{C}[1]{>{\centering\arraybackslash}m{#1}}
\usepackage{caption}
\usepackage{hyperref}
\usepackage{fancyhdr}

\usepackage{titling}

\usepackage{amsmath,amssymb}
\usepackage{slashed}
\usepackage{xcolor} 
\usepackage{graphicx}
\usepackage{dcolumn} 
\usepackage{bm} 
\usepackage{epsfig}
\usepackage{epstopdf}
\usepackage{grffile}
\usepackage{color}
\usepackage{colordvi}
\usepackage{amsmath,amssymb}
\usepackage{rotating}
\usepackage{lscape}
\usepackage{cite}
\usepackage{float}
\usepackage{hyperref}
\usepackage{url}
\usepackage{graphicx}
\usepackage{color}
\usepackage{amssymb}
\usepackage{amsmath}
\usepackage{mathtools}
\usepackage{bm}	
\usepackage{flexisym}
\usepackage{amssymb}
\usepackage{physics}
\usepackage{mathrsfs}
\usepackage{simplewick}
\usepackage{tikz}
\usetikzlibrary{arrows,shapes}
\usetikzlibrary{trees}
\usetikzlibrary{matrix,arrows} 			
\usetikzlibrary{positioning}			
\usetikzlibrary{calc,through}			
\usetikzlibrary{decorations.pathreplacing}  
\usepackage{pgffor}						
\usetikzlibrary{decorations.pathmorphing}
\usetikzlibrary{decorations.markings}
\tikzset{
	vector/.style={decorate, decoration={snake}, draw},
	provector/.style={decorate, decoration={snake,amplitude=2.5pt}, draw},
	antivector/.style={decorate, decoration={snake,amplitude=-2.5pt}, draw},
	fermion/.style={draw=black, postaction={decorate},
		decoration={markings,mark=at position .55 with {\arrow[draw=black]{>}}}},
	fermionbar/.style={draw=black, postaction={decorate},
		decoration={markings,mark=at position .55 with {\arrow[draw=black]{<}}}},
	fermionnoarrow/.style={draw=black},
	gluon/.style={decorate, draw=black,
		decoration={coil,amplitude=4pt, segment length=5pt}},
	scalar/.style={dashed,draw=black, postaction={decorate},
		decoration={markings,mark=at position .55 with {\arrow[draw=black]{>}}}},
	scalarbar/.style={dashed,draw=black, postaction={decorate},
		decoration={markings,mark=at position .55 with {\arrow[draw=black]{<}}}},
	scalarnoarrow/.style={dashed,draw=black},
	electron/.style={draw=black, postaction={decorate},
		decoration={markings,mark=at position .55 with {\arrow[draw=black]{>}}}},
	bigvector/.style={decorate, decoration={snake,amplitude=4pt}, draw},
}
\tikzstyle{block} = [draw, rectangle, minimum height=3em, minimum width=6em]
\usepackage{hyperref}

\usepackage{titling}
\usepackage{subcaption}
\newcommand{\subtitle}[1]{%
	\posttitle{%
		\par\end{center}
	\begin{center}\large#1\end{center}
	\vskip0.5em}%
}

\def\Re{{\cal R \mskip-4mu \lower.1ex \hbox{\it e}\,}}
\def\Im{{\cal I \mskip-5mu \lower.1ex \hbox{\it m}\,}}

\def\tev{\,{\ifmmode\mathrm {TeV}\else TeV\fi}}
\def\gev{\,{\ifmmode\mathrm {GeV}\else GeV\fi}}
\def\mev{\,{\ifmmode\mathrm {MeV}\else MeV\fi}}

\begin{document}

\begin{center}

\vspace*{15mm}
\vspace{1cm}
{\Large \bf Prospect study for measurement of $Hb\bar{b}$ coupling  at the LHC and FCC-hh}

\vspace{1cm}

{\bf Zahra Abdy$^{1}$  and Mojtaba Mohammadi Najafabadi$^{2,3}$ }

 \vspace*{0.5cm}

{\small\sl 
$^{1}$ AGH University of Science and Technology, Faculty of Physics and
Applied Computer Science, Al. Mickiewicza 30, 30-055 Krakow, Poland \\
$^{2}$Experimental Physics Department, CERN, 1211 Geneva 23, Switzerland  \\
$^{3}$ School of Particles and Accelerators, Institute for Research in Fundamental Sciences (IPM) P.O. Box 19395-5531, Tehran, Iran } \vspace*{.2cm}
\end{center}

\vspace*{10mm}

%
%
\begin{abstract}\label{abstract}

This paper employs the $H+b+\text{jets}$ signature in proton-proton collisions to explore the  structure of the 
$Hb\bar{b}$ couplings.The focus of the analysis lies in the decay of the Higgs boson into a photon pair, 
taking into account both reducible and
 irreducible backgrounds and a realistic simulation of the detector effects. 
 To enhance the discrimination between signal and background, 
a multivariate analysis is employed to analyse the kinematic variables and optimise the signal-to-background ratio. 
 The results indicate that the $H+b+\text{jets}$ process can significantly 
 contribute to the precise measurement of CP-even and CP-odd couplings 
 between the bottom quark and the Higgs boson at the LHC and FCC-hh.
Finally, a novel asymmetry is introduced for the purpose of probing CP violation within the 
$Hb\bar{b}$ coupling, formulated exclusively based on lab-frame momenta. 
\end{abstract}
\vspace{2cm}
{\bf Keywords}: {\small Higgs boson, bottom quark, CP violation, hadron colliders.}

\newpage

\section{Introduction }

The observation of the $H\rightarrow b\bar{b}$
decay mode by the ATLAS and CMS experiments has been made through  
the Higgs boson production in association with a massive vector boson $V$ ($V = W,Z$) processes  \cite{ATLAS:2018kot, CMS:2018nsn}. 
Both the ATLAS and CMS analyses rely on the leptonic decays of the vector boson ($W^{\pm}$ and $Z$) for triggering the events 
and to suppress the QCD multi-jet background.
The most recent measurements for observation of the $H\rightarrow b\bar{b}$ via $VH$ processes are presented 
using the whole Run 2 data collected by the ATLAS experiment in proton-proton
collisions with an integrated luminosity of 139 fb$^{-1}$ at $\sqrt{s} = 13$ TeV. 
For a Higgs boson with 125 GeV produced in either $HZ$ or $HW$ channel, an observed
significance of 6.7 standard deviations is obtained from the ATLAS experiment and the measured signal strength relative to the 
prediction of SM is found to be  \cite{ATLAS:2020fcp}:
\begin{eqnarray}
\mu_{bb} = 1.02 \pm 0.12 ~\text{(stat.)} \pm 0.14 ~\text{(syst.)},
\end{eqnarray}
Similar measurement by the CMS experiment is available with a combination of
Run 1 data (7 TeV and 8 TeV ) and part of Run 2 (2017 data corresponding to an integrated luminosity of 41.3 fb$^{-1}$).
The observed significance of 5.6 standard deviations and the measured signal strength from the combination $ZH$ and $WH$ processes is \cite{CMS:2018nsn}:
\begin{eqnarray}
 \mu_{bb}= 1.04 \pm 0.2 ~\text{(stat.} \oplus \text{syst.)},
\end{eqnarray}
As can be seen, the overall uncertainty is around $20\%$ in both measurements from the ATLAS and CMS experiments 
and the results are in agreement with the Standard Model (SM) prediction within the uncertainties. 
In the future, at the HL-LHC, an improvement of $10\%$ in tagging the b-quark jets efficiency is expected 
which leads to a relative improvement in the uncertainty of signal strength of up to $4.7\%$ \cite{CMS:2018qgz}.\\
A global fit of the Higgs boson couplings has been conducted in Ref.\cite{hbbnew} utilizing the comprehensive Higgs datasets collected at the LHC, 
encompassing integrated luminosities per experiment of approximately 5 fb$^{-1}$ at 7 TeV, 20 fb$^{-1}$ at 8 TeV, and up to 139 fb$^{-1}$ at 13 TeV. 
This analysis included the exploration of CP-even and CP-odd couplings of the Higgs boson to bottom quarks.\\
To probe the Higgs boson coupling with the SM particles and find any deviation from the SM predictions,
the $\kappa-$framework is used \cite{LHCHiggsCrossSectionWorkingGroup:2013rie}. In this framework, possible deviation
for the Higgs-bottom quark coupling is defined by $\kappa_{b}^{2} = \Gamma_{H\rightarrow b\bar{b}}/ \Gamma^{\rm SM}_{H\rightarrow b\bar{b}}$. 
The current measurement of $\kappa_{b}$ at $68\%$ confidence level (CL), obtained from a general fit to the Higgs boson couplings,
is $0.99^{+0.17}_{-0.16}$ \cite{CMS:2022dwd}. At HL-LHC, assuming similar systematic uncertainties to the Run 2 of the LHC, 
$\kappa_{b}$ is expected to be measured with an uncertainty of $\pm 0.042$ at $68\%$ CL \cite{Cepeda:2019klc}.\\
Except for some anomalies which are listed in Ref.\cite{Crivellin:2023gky} and references therein, all the experimental results obtained by the LHC 
experiments are consistent with the SM expectations within the uncertainties. In particular, the measured Higgs boson 
couplings with the SM fields are in agreement within the total uncertainties \cite{CMS:2022dwd}.
Therefore, any new degree of freedom could be foreseen to be well separated in mass from the SM particles \cite{r3,r4}.
As there are several beyond SM theories, with similar experimental signatures in some cases,
an easy and efficient way to search for new physics effects is to rely on a model independent way. 
In a model independent approach, the impacts of new physics could show up 
in an Effective Field Theory (EFT) extension of the SM which consists of
an infinite series of effective operators with higher dimensions \cite{r50,r51,r52,r53,r54,r55,r56,r566}.
The leading contributions to the effective extension of the SM originate from
the dimension 6 operators that is based on a non-redundant and complete operator basis \cite{r57,r58}.
Not only these operators can modify the signal strengths but also  
the differential distributions may be affected because of the presence of new vertex structures.
Several effective operators are
contributing to the Higgs boson couplings with the SM fields which persuade
us to pay attention to all possible Higgs involved processes at the colliders.
Probing processes where Higgs boson is present 
such as Higgs boson associated production, H+jets, and various
processes where Higgs boson is off-shell provides useful information of the new higher order effective couplings. 
There are already several studies for investigating the EFT extension of the SM in the 
Higgs boson sector \cite{r14, r15,r16,r17,r18,r19,r20,r21,r22,r23,r24,r25,r26,r27,r28,r29,r30,r1007,Khanpour:2017inb, Haghighat:2023bgi, r400, r500}.\\
Higgs boson production in association with a pair of $b\bar{b}$ at hadron colliders ($pp \rightarrow H+b+\bar{b}$),
has received attention as a direct way to probe both the CP-even and -odd structures of the 
bottom quark Yukawa coupling \cite{Grojean:2020ech, Konar:2021nkk}. 
In Ref.\cite{Grojean:2020ech}, using the kinematic shapes of $\gamma\gamma b\bar{b}$ 
final state in Boosted Decision Trees, a specific way to achieve  
a strong sensitivity to bottom quark Yukawa coupling is presented. 
The analysis has been performed at HL-LHC and Future Circular Collider proton-proton (FCC-hh)
considering the main sources of background processes. 
In conjunction with the HL-LHC, 
anticipated to operate at a center-of-mass energy of 14 TeV and an integrated luminosity of 3 ab$^{-1}$, 
the analysis of Ref.\cite{Grojean:2020ech} has been extended to the forthcoming FCC-hh facility. 
This future endeavor is poised to operate at an even higher center-of-mass energy,  
100 TeV, coupled with a substantially increased integrated luminosity of 15 ab$^{-1}$ \cite{fcchh}. 
The inclusion of FCC-hh as the succeeding flagship hadron collider initiative for the 
CERN, as indicated in the updated European strategy report, 
underscores its pivotal role in advancing the frontier of particle physics research \cite{Adolphsen:2022ibf}.\\
In Ref. \cite{Konar:2021nkk}, both a cut-based analysis and a gradient
 boost algorithm have been exploited 
to probe bottom quark Yukawa coupling
through $Hb\bar{b}$ channel with $4b$ final state. This analysis
only focuses on probing the CP-even structure of the $Hb\bar{b}$ coupling.\\
 The primary objective of this research paper is to investigate the $H+b+(\text{jet})$ process and 
 its potential to extract the CP-even and -odd components of the bottom quark Yukawa coupling 
 within the framework of Effective Field Theory (EFT). The approach employed for this purpose involves utilizing Machine Learning techniques.
The $H+b+(\text{jet})$ process under study allows for the inclusion of either a light flavour jet or a b-quark jet. 
The rationale behind exploring this process is two-fold. Firstly, by examining the capabilities of 
the $H+b+(\text{jet})$ process to probe the structure of $Hb\bar{b}$ couplings, a comparative analysis 
can be performed in relation to other relevant processes. Secondly, the research delves into the 
efficacy of a multivariate technique that leverages the shapes of the final state to discriminate between the signal and the main SM background processes.
Through this investigation, we aim to gain valuable insights into the underlying physics of the $H+b+(\text{jet})$ process 
and its significance in probing both CP-odd and CP-even $Hb\bar{b}$ couplings. Additionally, we seek to assess the strength of the 
Machine Learning-based multivariate approach in effectively distinguishing the signal from the SM background processes.\\
This paper is organised as follows. The theoretical framework of the 
$Hb\bar{b}$ effective coupling is described in section \ref{sec1}. 
In section \ref{sec12}, we discuss the Higgs production 
in  association with a bottom quark and additional jet. 
In section \ref{sec2}, the analysis strategy and details of event simulation are explained. 
The results of the analysis, exemplified for the HL-LHC and for the FCC-hh, are given in section \ref{sec3}. 
An asymmetry-like observable is introduced in section \ref{seccp} to 
explore the CP-violating term of the $Hb\bar{b}$ coupling.
Finally, we summarise and conclude in section \ref{sec4}.

\section{Theoretical framework}
\label{sec1}

As referenced in the preceding section, this study has been conducted within the framework of the EFT, 
where manifestations of new physics are anticipated to emerge through novel interactions among the SM fields. 
In this context, the introduction of new couplings is suppressed by inverse powers of $\Lambda$, which serves as the characteristic 
scale representing physics beyond the SM.
In the EFT,  the effects of  heavy new degrees of freedom are integrated out and the SM gauge symmetries, Lorentz invariance
and lepton and baryon number conservation are respected.
The new physics effects are parameterised by higher dimension
operators with not-known Wilson coefficients and the main contributions to the observable
come from dimension-six operators. In this work we rely on the Higgs characterisation
model \cite{hcm} based on an EFT approach where the $Hf\bar{f}$ interaction has the following 
form:
\begin{eqnarray} \label{effL}
\mathcal{L}_{Hf\bar{f}} = - \sum_{f = e,\mu,\tau,u,d,c,s,b,t} \frac{y_{f}^{\rm SM}}{\sqrt{2}}\bar{f}(c_{f} + i\tilde{c}_{f}\gamma_{5})fH,
\end{eqnarray}
where the Higgs boson field is denoted by $H$ and $f$ is the fermion 
field. The SM Yukawa coupling of a fermion $f$ is shown by coupling $y_{f}^{\rm SM}$.
Modifications from the dimension six operators to the $Hf\bar{f}$ CP-even and CP-odd couplings
show up in $c_{f}$ and $\tilde{c}_{f}$ parameters, respectively. In the SM, 
$c_{f} = 1.0$ and $\tilde{c}_{f}  = 0.0$ and $ y_{f}^{\rm SM}/\sqrt{2} = m_{f}/v$, where $v$
is the vacuum expectation value.  Both $c_{f}$ and $\tilde{c}_{f}$ for the top and bottom quarks can be indirectly studied 
through the measured Higgs boson production cross section via gluon-gluon fusion ($\sigma(gg\rightarrow H)$) and the 
decay width of the Higgs boson into two photons ($\Gamma(H\rightarrow \gamma\gamma)$). The modifications that 
$\Gamma(H\rightarrow \gamma\gamma)$ and $ \sigma(gg\rightarrow H)$ receive only from third quark generation
are \cite{r15}:
\begin{eqnarray}\label{kappaG}
 \kappa^{2}_{\gamma} &=& 0.08c^{2}_{t} + 0.18\tilde{c}^{2}_{t} + 4\times 10^{-5}\times(c^{2}_{b} + \tilde{c}^{2}_{b} )- 0.002\times c_{t}c_{b}-0.004 \tilde{c}_{t} \tilde{c}_{b}, \\ \nonumber
\kappa^{2}_{g} &=& 1.11c^{2}_{t} + 2.56\tilde{c}^{2}_{t} + 0.01\times(c^{2}_{b} + \tilde{c}^{2}_{b} )- 0.12\times c_{t}c_{b}-0.2 \tilde{c}_{t} \tilde{c}_{b}, 
\end{eqnarray}
where $\kappa_{g} = \sigma(gg\rightarrow H)/\sigma_{\rm SM}(gg\rightarrow H)$ and $\kappa_{\gamma} = \Gamma(H\rightarrow \gamma\gamma)/\Gamma_{\rm SM}(H\rightarrow \gamma\gamma)$.
As can be seen, the gluon fusion cross section can deviate significantly from its SM prediction 
even with minor deviations of $c_{t}$ and $\tilde{c}_{t}$ from their SM values. 
However, the impact of $c_{t}$ and $\tilde{c}_{t}$ on $\kappa_{\gamma}$ is less pronounced. 
When compared to the top quark coupling, variations in the coupling modifiers of the bottom quark ($c_{b}$ and $\tilde{c}_{b}$) exhibit much smaller magnitudes
in $\kappa_{g}$ and $\kappa_{\gamma}$. As discussed in Ref.\cite{r15}, from Eq.\ref{kappaG}, 
the gluon fusion cross section exhibits an enhancement of approximately $26\%$ when considering a negative value for $c_{b}$ parameter falling within $2\sigma$ allowed region, specifically $-1.23 \leq c_{b} \leq -1.08$. Meanwhile, the di-photon decay width experiences a reduction of around $3\%$ within this specified region.
On the other hand, the impact of $\tilde{c}_{b}$ on $\kappa_{g}$ and $\kappa_{\gamma}$ remains 
at the sub-percent level and loose limits on $\tilde{c}_{b}$ could be derived from $\kappa_{g}$ and $\kappa_{\gamma}$ \cite{r15}.\\

In addition to the gluon fusion Higgs boson cross section and di-photon decay width of the Higgs boson,
the CP-odd component of the $Hb\bar{b}$ coupling, $\tilde{c}_{b}$, can be constrained using the electron electric dipole moment (EDM).
This contribution to the EDM of the electron, $d_{e}$, occurs through loop processes. 
Consequently, the value of $\tilde{c}_{b}$ can be indirectly constrained by considering the existing experimental limit on $d_{e}$ \cite{r500}.
The ACME collaboration has established an experimental limit on the electron EDM at a $90\%$ confidence level to be 
$|d_{e}| \leq 1.1 \times 10^{-29}$ e.cm \cite{amce}.
Using this bound on the electron EDM, the constraint on $\tilde{c}_{b}$ at a $90\%$ CL is found to be $\tilde{c}_{b} < 0.26$. 
It is important to note that this limit is derived under the assumption that there are no deviations from the SM in
 the $He\bar{e}$ coupling and that no cancellations occur with other contributing mechanisms.

As a direct way to probe the $Hb\bar{b}$ interaction, the analysis of $H+b+\bar{b}$ production at the LHC and FCC-hh is expected to  provide
the following constraints at $1\sigma$ \cite{Grojean:2020ech}:
\begin{eqnarray}
&&c_{b} \in [-0.99,-0.82] \cup [0.84,1.14], \text{ at HL-LHC with  6 ab}^{-1} \\ \nonumber
&&c_{b} \in [0.99,1.01], \text{at FCC-hh with  of 30 ab}^{-1}.
\end{eqnarray}
where the result of HL-LHC and FCC-hh are obtained with an integrated luminosity of 6 ab$^{-1}$ and
30 ab$^{-1}$, respectively.

\section{Higgs boson production associated with a bottom quark jet and an additional jet }
\label{sec12}

\begin{figure}[h] 
	\centering	
	\includegraphics[width=0.5\textwidth]{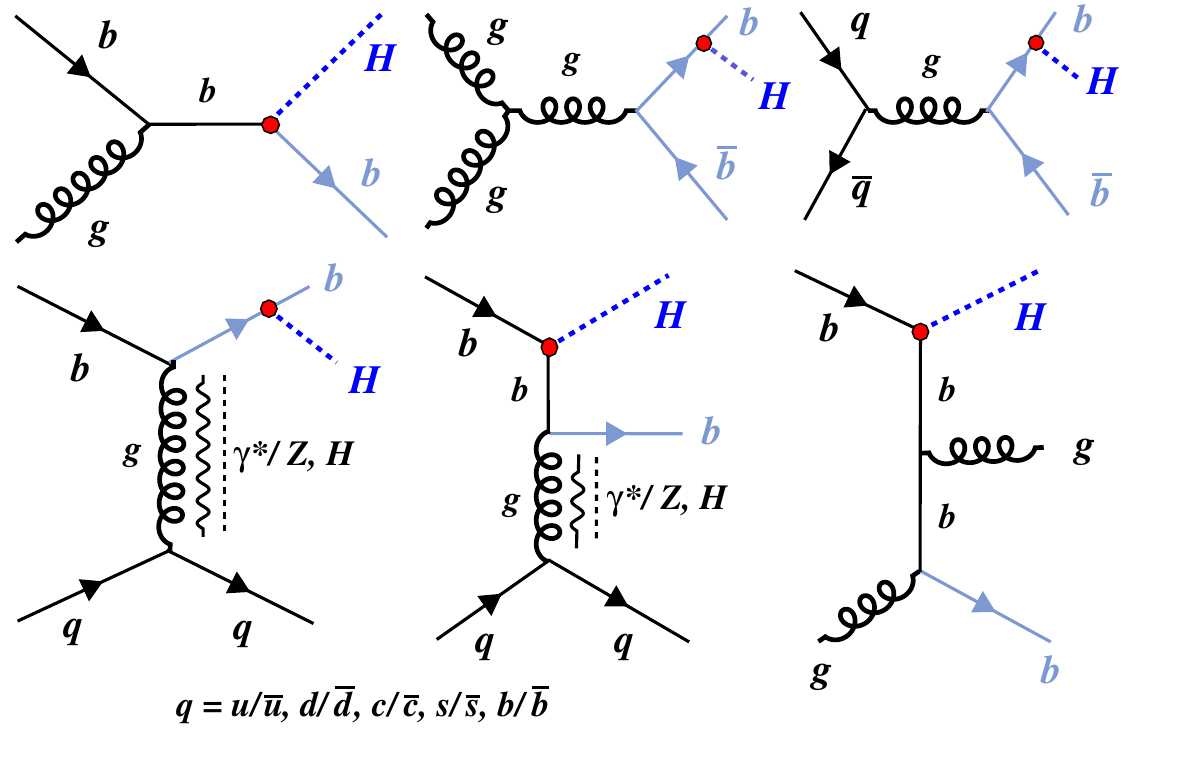}
	\caption{ Representative Feynman diagrams for production of Higgs boson in association with 
	a bottom quark and an additional parton at leading order.
	The filled red circle presents the effective coupling of the $Hb\bar{b}$. }\label{fig:Feynman}
\end{figure}

In this section, we describe the production of a Higgs boson associated with a b-quark jet and additional jet in proton-proton collisions 
which occur at the LHC and FCC-hh.
In the SM,  the $H+b+$(jet) production proceeds through
the gluon-$b$ quark interactions, quark-(anti)quark annihilation, and gluon-gluon fusion. 
The representative Feynman diagrams
are shown in Fig. \ref{fig:Feynman}. 
The filled circles represent the vertices in the diagram that undergo modifications 
due to the effective interaction of $Hb\bar{b}$ introduced in Eq.\ref{effL}.

The cross section for $H+b$ processes, involving the production of a Higgs boson and a bottom quark, 
is of the order of $\mathcal{O}(\alpha_s(y_{b}^{\rm SM})^2 )$, where $y_{b}^{\rm SM}$ represents the 
Yukawa coupling strength of the b-quark and $\alpha_s$ denotes the strong coupling constant.
On the other hand, the higher-order processes that include an additional jet have a cross section 
of $\mathcal{O}(\alpha_s^2(y_{b}^{\rm SM})^2 )$. 
These processes involve the emission of an extra jet alongside the Higgs boson and bottom quark, 
resulting in a higher suppression factor due to the additional power of 
the strong coupling constant.

To gain insights into the impact of varying the coupling modifiers at the LHC, we examine the changes in the signal 
cross section concerning the SM when $c_b$ is altered by approximately $10\%$ from their SM values. For $c_b=1.1$, 
the relative change in the cross section is approximately $18\%$, whereas for $c_b=0.9$, $|\Delta\sigma|/\sigma_{\rm SM}=17\%$.
Additionally, the effect of changing $\tilde{c}_b$ within $\pm 20\%$ and $\pm 40\%$ of the SM value, 
results in relative changes of approximately $3\%$ and $21\%$, respectively. 

To assess the effects of the $Hb\bar{b}$ effective couplings on $Hb(H\bar{b})$ productions in association with a jet, 
the {\tt MadGraph5\_aMC@NLO} package (version 3.5.1) \cite{Alwall:2011uj, Alwall:2014bza,Alwall:2014hca} is utilised. 
This package allows for the calculation of cross-sections and event generation for various processes. 
In this case, the effective Lagrangian, as introduced in Eq.\ref{effL}, is implemented in the {\tt FeynRule} program. 
The resulting model, known as the Universal FeynRules Output (UFO) model \cite{Alloul:2013bka}, is then provided as input to the {\tt MadGraph5\_aMC@NLO} program.
The production of $Hb/H\bar{b}$ with an additional parton in the final state is considered at leading order, employing relevant matrix elements. 
The events with zero and one additional parton are combined using the MLM matching scheme \cite{mlm}, which ensures a consistent description of 
the production process.

In this study, we aim to determine a realistic sensitivity of the $H+b$+jet process to general $Hb\bar{b}$ coupling, 
specifically focusing on the $c_{b}$ and $\tilde{c}_{b}$. To achieve this, we carry out the analysis using the Higgs 
decay channel to two photons, which is known for its well-reconstructed and rather clean signature. 
This choice allows us to well estimate the background processes and provides valuable insights into the impact of different 
contributions on $c_{b}$ and $\tilde{c}_{b}$. 
With Higgs boson decaying into two photons, the final state consists of two photons and at least one jet
from which at least one is originating from the hadronization of a b-quark.

The analysis considers several dominant background processes that contribute to the signal. 
These background processes, which need to be carefully taken into account, include:
\begin{itemize}
\item{SM production of $H+b$+jet (merged $H+b$ and $H+b$+jet using MLM prescription), where Higgs boson decays into $\gamma+\gamma$ and
jet can be a light or a heavy flavour jet.}
\item{Higgs boson production from gluon-gluon fusion process: $pp \rightarrow H \rightarrow \gamma+\gamma$.}
\item{Higgs boson production in association with a vector boson: $pp \rightarrow H+V\rightarrow \gamma+\gamma+ jets$, where $V = W^{\pm}$ and $Z$.}
\item{Di-photon production associated with b-quark jets ($\text{b-jets}+\gamma\gamma+\text{jets}$). }
\item{Di-photon production associated with c-quark jets ($\text{c-jets}+\gamma\gamma+\text{jets}$). This background arises from the production of a pair of photon, not coming from Higgs boson,
 accompanied by c-quark jets, where the c-jets are mis-tagged as b-jets.}
\item{Di-photon production associated with light flavour jets where light flavour jets are misidentified as b-jets. }
\end{itemize}

To ensure sufficient statistics and obtain a more precise estimate of their contributions, the di-photon production 
associated with different types of jets, including b-quark jets, c-quark jets, and light flavour (non-bottom, non-charm) quark jets, 
is generated separately in the analysis. By generating these processes independently, it becomes possible to have an 
adequate number of events for each specific jet flavour and accurately estimate their individual contributions to the di-photon final state. 
This approach allows for a more comprehensive understanding of the impact of different jet flavors on the di-photon signal and helps
 in properly accounting for their respective backgrounds in the analysis.

\section{Simulation and Analysis Strategy}
\label{sec2}

The SM background processes and signal events are generated using the {\tt MadGraph5\_aMC@NLO} event generator. 
Specifically, the Higgs boson production in association with a b-quark and a jet sample is obtained by merging the Higgs boson plus b-quark ($H+b$) 
sample and the Higgs boson plus b-quark plus jet ($H+b$+jet) sample using the MLM merging prescription.
By merging the samples, a comprehensive description of the Higgs boson production 
in association with a b-quark and a jet is achieved, allowing for accurate modelling and analysis of the signal and background processes.

After generating the samples, they undergo further simulation and modelling to account for various effects. 
First, the samples are passed through {\tt PYTHIA 8.3}  \cite{Sjostrand:2014zea}, which handles parton showering, hadronisation, and the decay of unstable particles. 
To account for the effects of the detector, the {\tt Delphes 3.5.0} package \cite{deFavereau:2013fsa} is utilised, which simulates both CMS detector phase II card and
the FCC-hh card. Delphes takes the generated particles as input and applies realistic detector response and resolution effects to the particles.
For jet reconstruction, the anti-$k_{\rm t}$ algorithm \cite{Cacciari:2008gp} with a cone size parameter $R=0.4$, implemented in the FastJet package \cite{Cacciari:2011ma}, is employed. 
This algorithm clusters particles into jets based on their proximity in the detector.
To identify and tag jets originating from b-quarks, b-tagging efficiency and misidentification rates are considered. 
The efficiency of b-tagging for a jet is $p_{\rm T}$ and $\eta$ dependent. 
Additionally, misidentification rates for charm-jets and light-flavour jets are taken into account.
For example, for the HL-LHC with CMS Phase II Delphes card, the b-tagging efficiency for a jet with $p_{\rm T} \in [40,50]$ and $|\eta| \leq 1.8$ is set to $66.6\%$. 
The misidentification rate for the charm-jets is $p_{\rm T}-\eta$ dependent and for the light-flavour jets is flat. 
For a charm-jet with $p_{\rm T} \in [40,50]$ and $|\eta| \leq 1.8$ the misidentification rate is
$18.8\%$ while it is $1.0\%$ for light-flavour jets independent of $p_{\rm T}$ and $\eta$. 
These rates reflect the likelihood of mistakenly tagging a charm-jet or a light-flavour jet as a b-jet.
By incorporating these simulation and modelling steps, the analysis aims to realistically account for detector effects, jet reconstruction, 
and the identification of b-jets, charm-jets, and light-flavour jets, thereby providing a more accurate description of the experimental observables.

\subsection{Events Selection}
\label{evsel}

To identify signal events, specific criteria are applied. The event selection requires the presence of exactly 
two isolated photons. These photons must have a transverse momentum greater than or equal to 20 GeV 
and a pseudorapidity  within the range of $|\eta_{\gamma_{1,2}}| \leq 3.0$. 
An isolated photon is defined as one that exhibits minimal activity in its vicinity, reducing the probability of originating from a jet. 
A small value of the isolation variable indicates a high degree of isolation, 
implying that the photon is more likely to be a genuine rather than originating from a jet.
The isolation of a photon, $I_{\gamma}$ , is quantified using an isolation variable, which is calculated as the ratio 
of the sum of transverse momenta ($p_{\rm T}$) 
of other particles around the photon to the transverse momentum of the photon itself. For both photons
the isolation variable is required to be less than $0.15$.

Additionally, each event must contain at least one jet, out of which at least one must be identified as b-jets using a b-tagging criterion. 
The jets (b-jets) are required to have a $p_{\rm T}$ greater than or equal to 30 GeV and a pseudorapidity within the range of $|\eta| \leq 4.0 ~ (3.0)$.

To ensure that the selected objects are well-isolated, an angular separation criterion is applied. 
The angular separation between any two objects (photons or jets) is quantified using the variable
 $\Delta R_{i,j} = \sqrt{(\eta_{i} - \eta_{j})^{2} + (\phi_{i} - \phi_{j})^{2}}$, where $\phi$ represents the azimuthal 
 angle and $\eta$ represents the pseudorapidity. 
 The requirement is set at $\Delta R_{i,j} \geq 0.4$ for all possible combinations of objects (photon-photon, photon-jet, and jet-jet). 
 This criterion helps to ensure that the selected objects are sufficiently separated from each other in the detector.

By applying these selection criteria, the analysis aims to isolate signal events that satisfy the specific kinematic and isolation requirements, 
ensuring the quality and reliability of the data used for further analysis. 
Table \ref{effc} displays the efficiencies of the signal  when the coupling  
$c_{b}$ is set to 1.05  and  the coupling  $\tilde{c}_{b}$ is set to 0.0. It also includes the efficiencies of 
the main background processes after applying the cuts. The efficiencies represent the fraction of events that pass the selection criteria, for each process.

\begin{table}[htbp] 
\begin{center}
\begin{tabular}{|c|c|c|c|c|c|}
\hline
     & Signal  & SM $H+b+\text{jets}$ & b-jets+$\gamma\gamma$ & c-jets+$\gamma\gamma$ & light-jets+$\gamma\gamma$ \\ \hline
HL-LHC & $3.6\%$ & $3.4\%$         & $5.1\%$               & $0.25\%$              & $0.05\%$                  \\ \hline
FCC-hh & $4.3\%$ & $4.1\%$         & $2.7\%$               & $0.58\%$               & $0.09\%$                   \\ \hline
\end{tabular}
\end{center}
\caption{ The efficiencies of the signal with $c_{b} = 1.05, \tilde{c}_{b} = 0.0$ and for the main SM background processes after applying the selection criteria.} \label{effc}
\end{table}

The cross sections of other background processes such as $H, HZ$, and $HW$ after the selection are found to be
negligible in both HL-LHC and FCC-hh. 
At the end of this section, we address the potential impact of a background arising from the misidentification of jets as photons in multijet
and jets+$\gamma$ production. 
This occurs when jets contain neutral pions that decay into two photons, resulting in overlapping photon showers that appear as a single photon in the detector.
To mitigate the QCD multijet background, it is crucial to develop methods to identify and reject jets mimicking photons. 
This background process has a significantly higher cross section compared to other backgrounds, with a difference 
spanning several orders of magnitude. However, by applying kinematic requirements (excluding criteria related to photons) 
and ensuring the presence of only  b-jets, the cross section of the multijet and jets+$\gamma$ background is reduced to approximately $10^{4}$  pb.
The probability of a jet being misidentified as a photon depends on the transverse momentum of the fake photon, typically 
ranging from $10^{-5}$ to $10^{-3}$ depending on fake photons $p_{\rm T}$. By imposing a requirement of two photons, 
the contribution of this background is effectively minimised to a low level.
While this analysis neglects the multijet background, it is important to emphasise that a dedicated and more realistic detector 
simulation is necessary to accurately estimate its potential contribution. Such a simulation should consider the specific 
characteristics of the experimental setup and incorporate a detailed representation of the detector response. 
Future studies should address this aspect to provide a more comprehensive assessment of the multijet background's impact.

\subsection{Multivariate analysis}

In this analysis, the applied cuts on individual variables are generally loose, meaning they do not significantly suppress a 
substantial fraction of background events while also reducing the signal events. 
To achieve a better discrimination between the signal and background processes a multivariate analysis is
used \cite{tmva1, tmva2, tmva3, tmva4, tmva5}.  In particular a  Boosted Decision Tree (BDT) is trained to increase the sensitivity.
All the backgrounds are taken into account during the BDT training, with each background process weighted accordingly. 
This helps in obtaining an effective separation of signal events from the background events.
To enhance sensitivity, separate analyses are conducted for the CP-even and CP-odd signals using 
distinct BDT test and training sets. This approach allows for a more targeted investigation of each signal, 
optimizing the discrimination power and enhancing the overall sensitivity of the analysis.
Distinct sets of variables for  CP-even ($c_{b}$) and CP-odd ($\tilde{c}_{b}$) signal processes are employed in the BDT
models for analyses at both the HL-LHC and FCC-hh. The included variables are as follows:
\begin{itemize}
	\item {$p_{T, \gamma_{1}} , p_{T, \gamma_{2}}:$ transverse momenta of first and second photon. }
	\item {$ p_{T,\rm b-jet}: $ transverse momentum of the most energetic b-jet.}
	\item{ $p_{ T,\gamma_1 \gamma_2}= |\vec{p}_{T,\gamma_1} + \vec{p}_{ T, \gamma_2} |$:  the magnitude of the vector sum 
	of the transverse momenta of the two photons.}
	\item {$ M_{\gamma_1 \gamma_2}$ : invariant mass of two  photons (reconstructed Higgs boson mass). }	
	\item {$ M_{\gamma_1, \gamma_2 , \rm b-jet}$:  the invariant mass of two photons and the leading b-jet system. }
	\item {$ M_{\gamma_1 ,\rm  b-jet}$:  the invariant mass of  leading photon and  leading b-jet system. }
	\item {$ M_{\gamma_2 , \rm b-jet}$:  the invariant mass of second photon and first leading b-jet system. }
	\item {$ \Delta R (\gamma_1 ,\gamma_2)$: the angular distance between the first and second photon. }
	\item{$ \cos(\gamma_1 ,\gamma_2)$: the cosine of angle between the two photons.}
	\item{$\Delta\phi(\gamma_{1,2},\rm b-jet)$: the azimuthal angle between the photons and the b-jet. These azimuthal angle variables are expected to be sensitive to 
	the CP-violating coupling \cite{cp1,cp2,cp3,cp4}. }
	\item{mean IP of the b-tagged jet defined as: \\
	\begin{eqnarray}\label{ippp}
	\frac{\sum_{\rm i=trk }^{N_{trk}}\sqrt{d_{0,i}^2 + d_{z,i}^2}}{N_{trk}}, 
	\end{eqnarray}
	where $d_{0,i}$ and $d_z,i$ are the transverse impact parameter and the longitudinal impact parameter values of $i$th track 
	 inside the b-tagged jet cone with $\Delta R = 0.4$. The total number of tracks inside the jet cone is $N_{trk}$. }
	\item{b-jet charge (${\mathbf  Q_{\rm b-jet}}$):  
	\begin{eqnarray}\label{qjjj}
	    \frac{\sum_{i} q_{i} \times p_{ T, i}}{p_{T, \rm b-jet}},
	\end{eqnarray}
	where the sum is over the particles inside the reconstructed b-jet cone, $q_{i}$ is the integer charge value of the observed color-neutral object, 
	 $p_{\rm T, i}$ is the magnitude of its transverse momentum w.r.t the beam axis, and the total transverse momentum of the b-jet is denoted 
	 by $p_{\rm T,b-jet}$. More details of the jet charge definition is found Ref.\cite{ref_jetcharge}.
	}
\end{itemize}

\begin{figure}[ht!]
\centering
	\includegraphics[width=0.45 \textwidth]{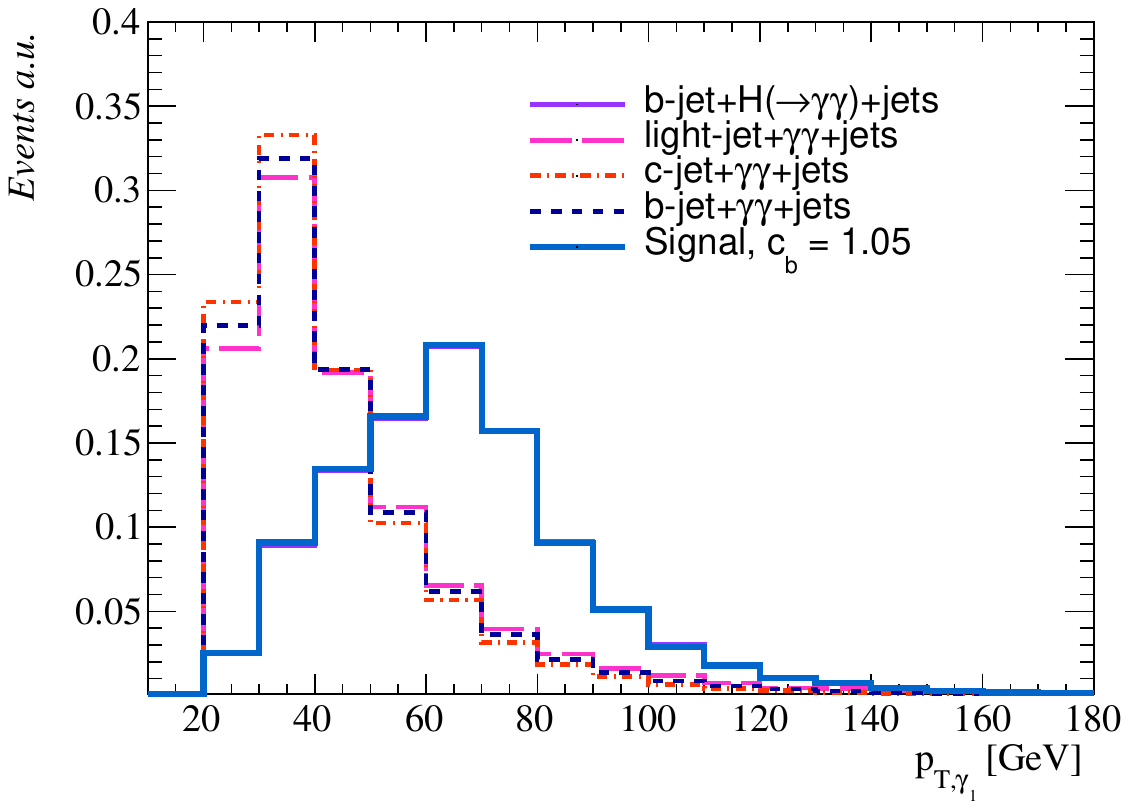}
	\includegraphics[width=0.45\textwidth]{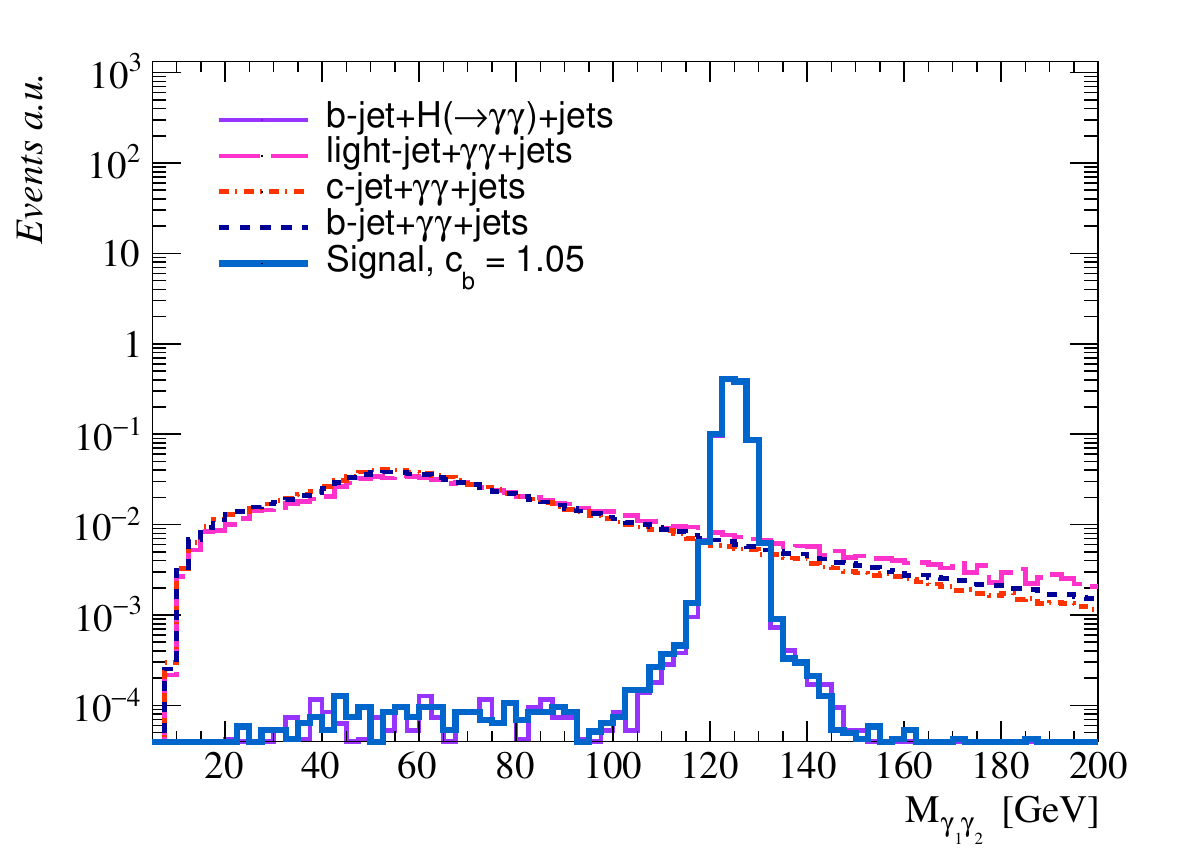}
	\includegraphics[width=0.45\textwidth]{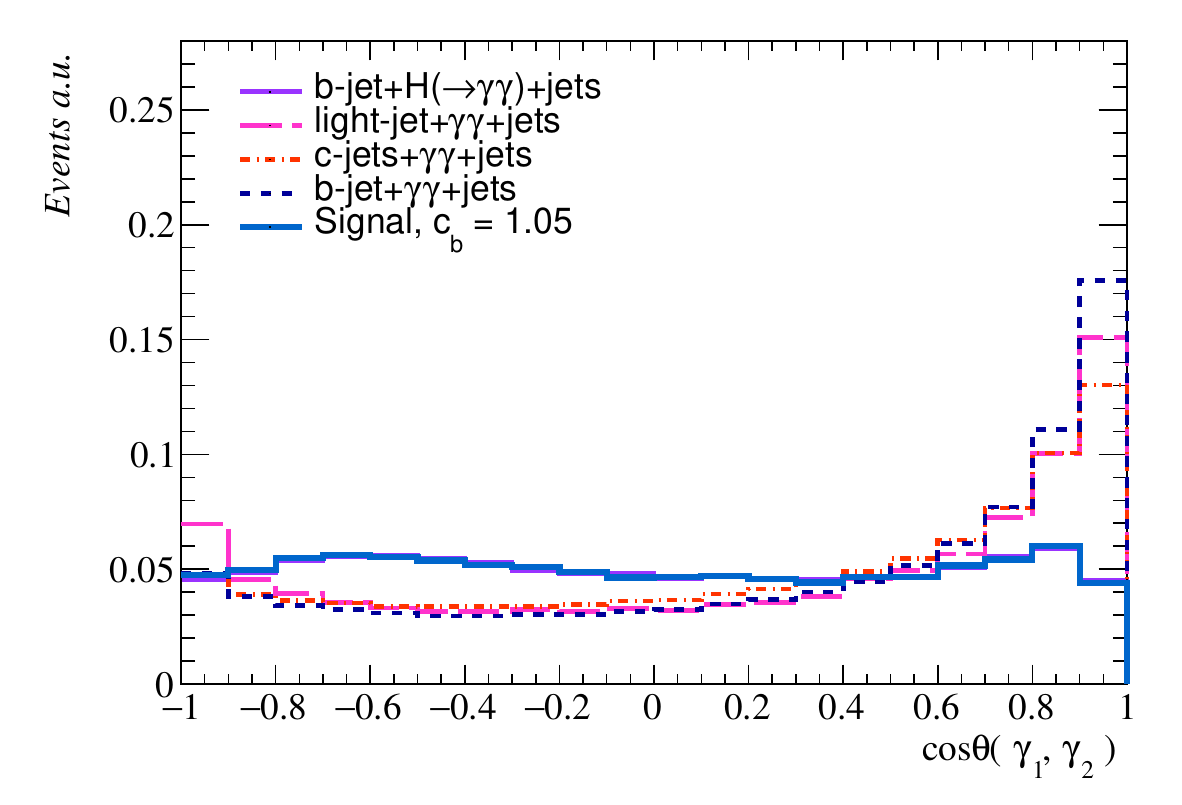}
	\includegraphics[width=0.45\textwidth]{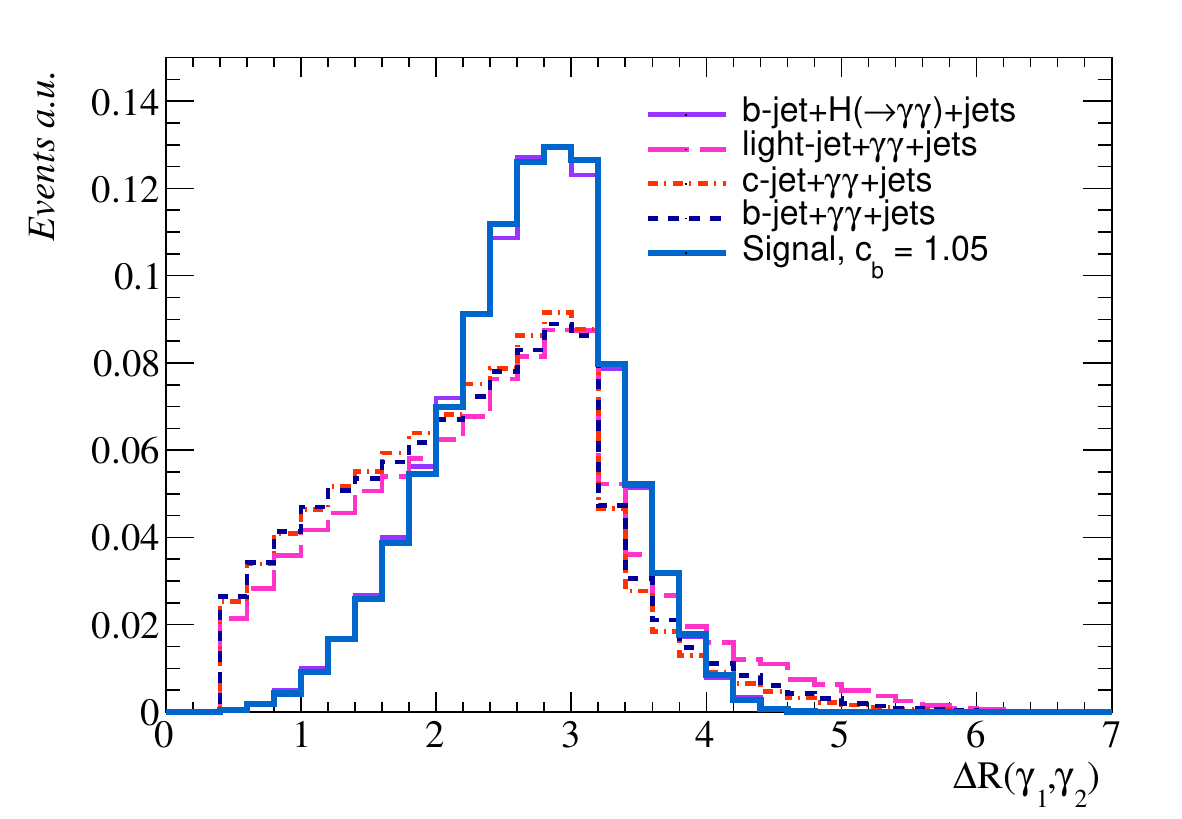}\\
	\caption{Normalised distributions for the signal with $c_{b} = 1.05, \tilde{c}_{b} = 0.0$ and the main background processes of some of the input
variables at the HL-LHC used in the BDT after the selection criteria.  In particular, first photon $p_{T}$,  
invariant mass of diphoton, cosine of the angle between two photons ($\cos(\gamma_1, \gamma_2)$)
, and  the angular distance between the di-photon system ($\Delta R(\gamma_1, \gamma_2)$).}\label{fig:BDTinputHLLHC1}
\end{figure}

\begin{figure}[ht!]
\centering
	\includegraphics[width=0.45 \textwidth]{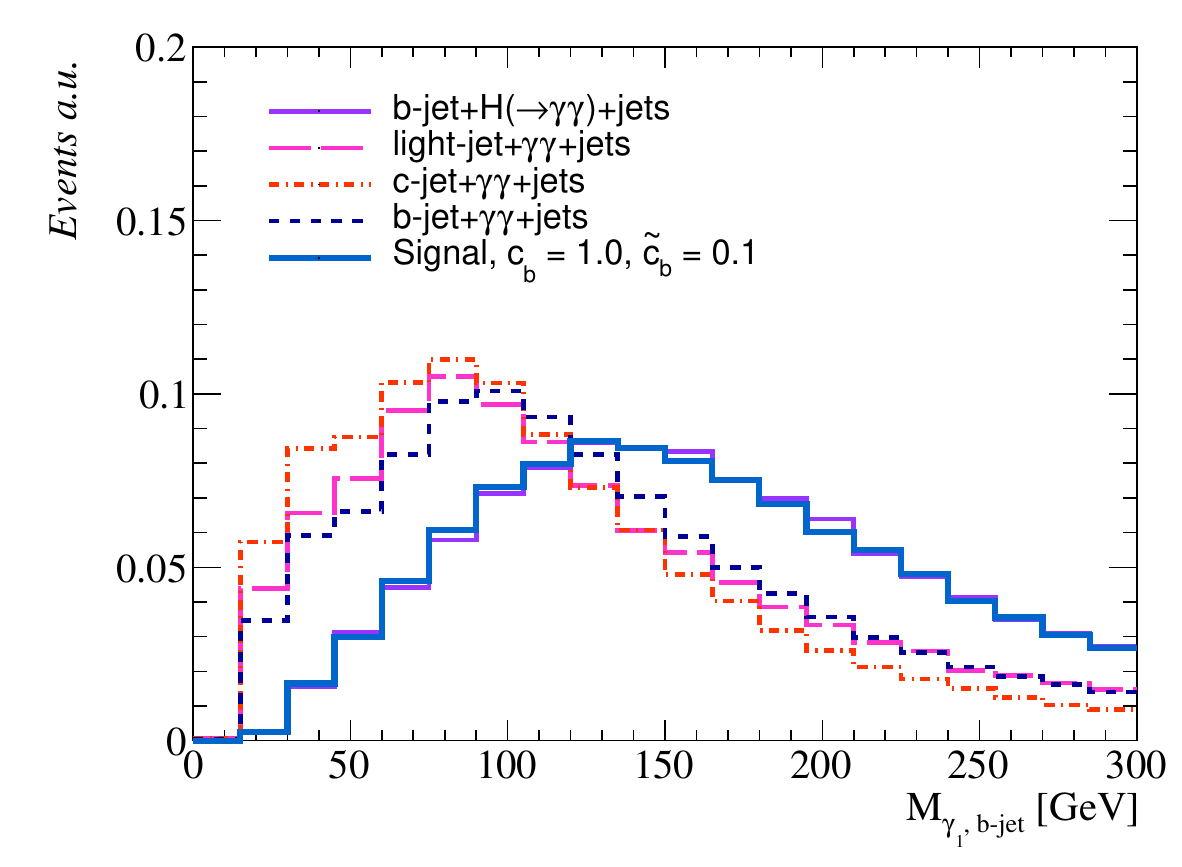}
	\includegraphics[width=0.45\textwidth]{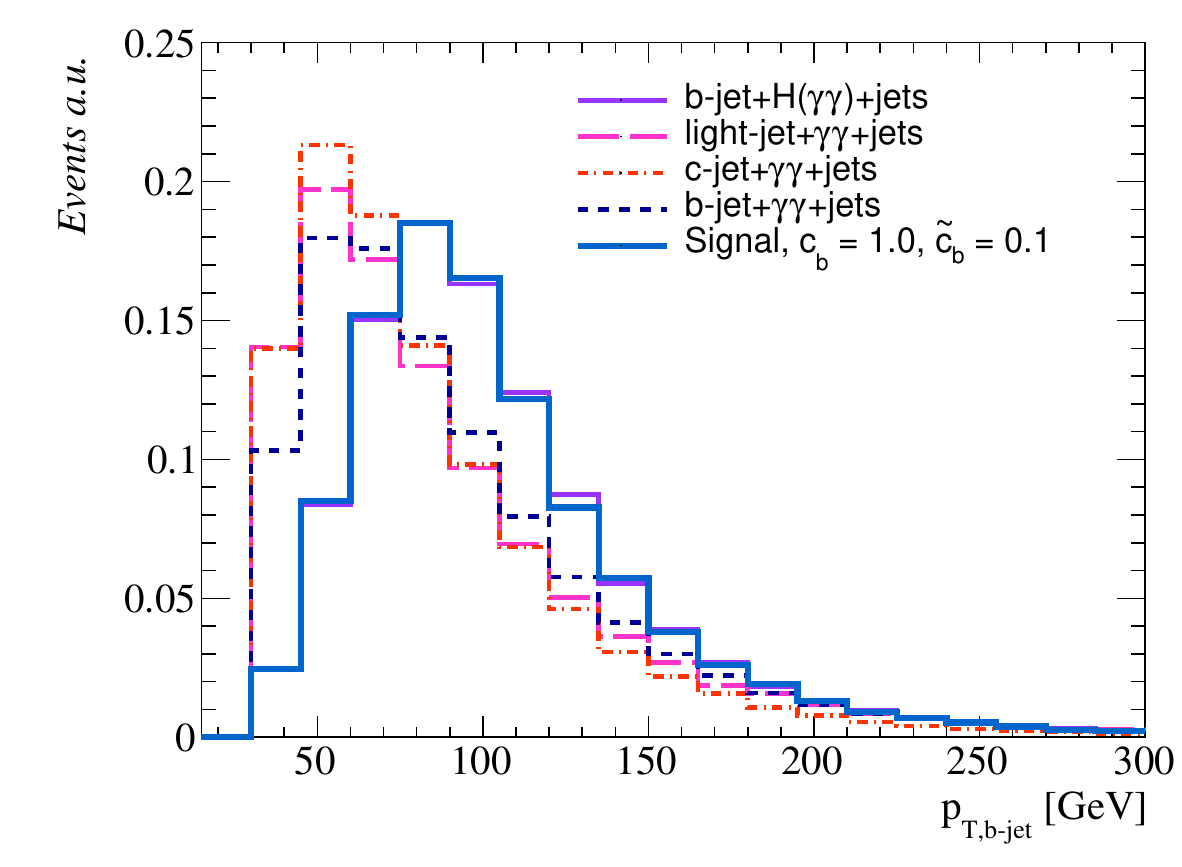}
	\includegraphics[width=0.45\textwidth]{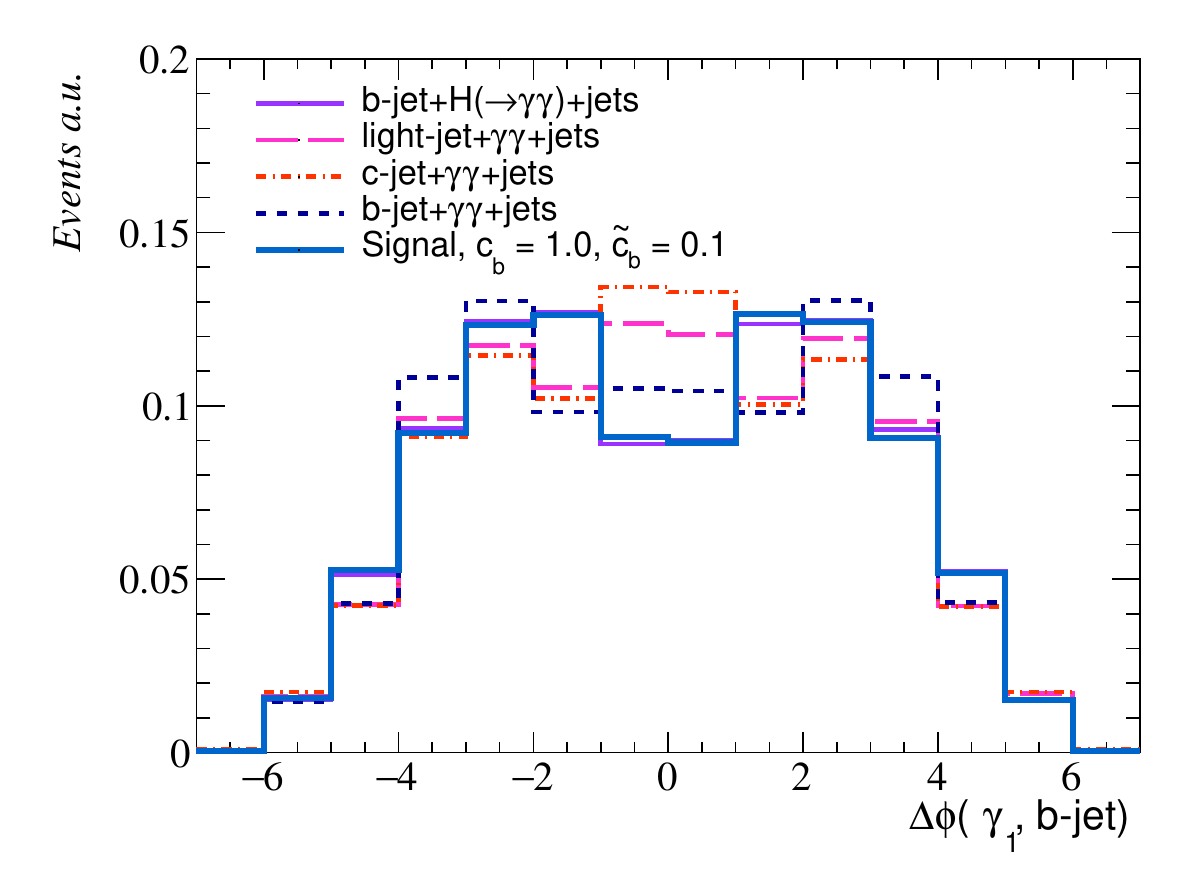}
	\includegraphics[width=0.45\textwidth]{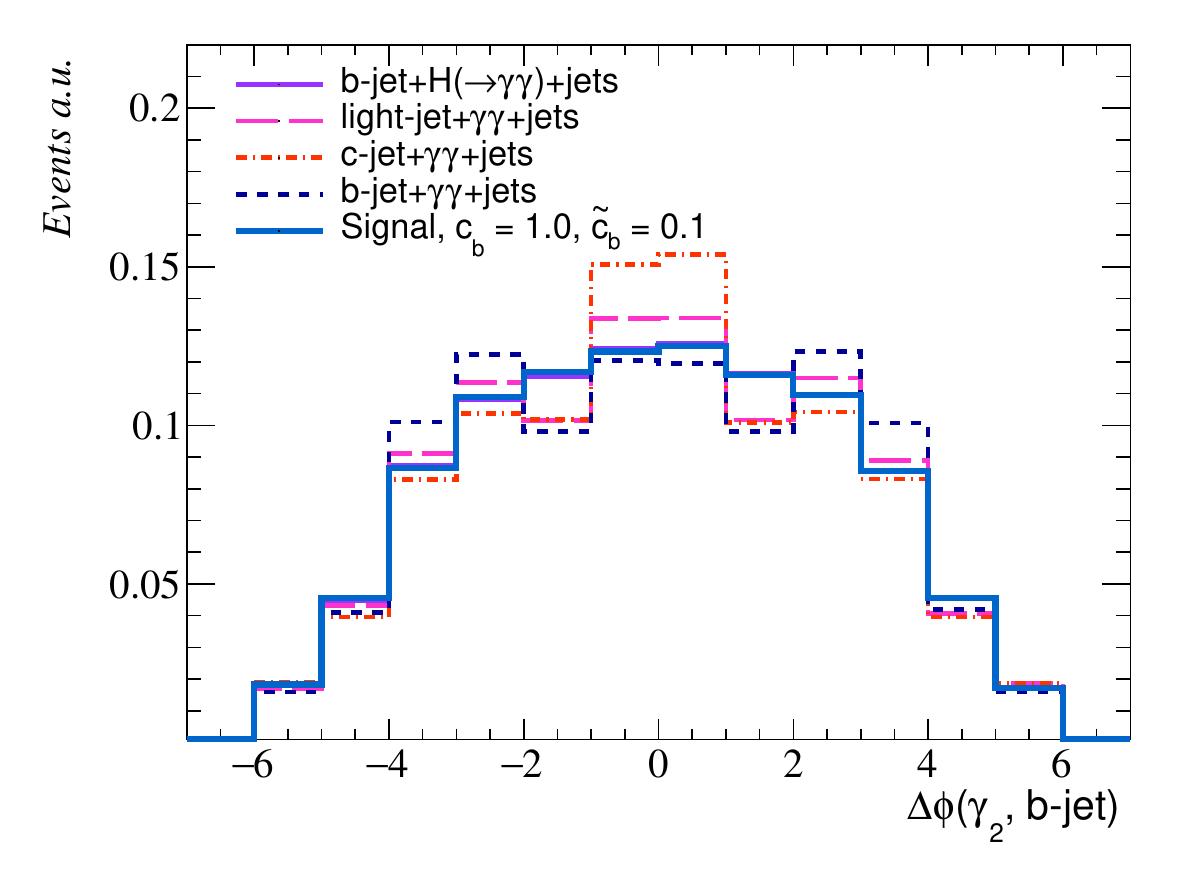}\\
	\caption{Normalised distributions for the CP-odd signal events with $c_{b} = 1.0, \tilde{c}_{b} = 0.1$ and the main background processes of some of the input
variables at the HL-LHC used in the BDT after the selection criteria.  The invariant mass of the leading photon and the leading jet system ($M_{\gamma_1,\rm b-jet}$), leading b-jet $p_{T}$
, and  the difference between the azimuthal angles of $\gamma_{1,2}$ and the leading b-jet ($\Delta\phi(\gamma_{1,2},\rm b-jet)$).}\label{fig:BDTinputHLLHC2}
\end{figure}

Figure \ref{fig:BDTinputHLLHC1} displays the normalised distributions of some the input variables for the CP-even signal. The distributions
for signal are for the case $c_{b} = 1.05, \tilde{c}_{b} = 0.0$.
In particular, first photon $p_{T}$,  
invariant mass of diphoton, cosine of the angle between two photons ($\cos(\gamma_1, \gamma_2)$), and  the angular distance between the di-photon system ($\Delta R(\gamma_1, \gamma_2)$) are depicted in Fig.\ref{fig:BDTinputHLLHC1}.
The normalised distributions of some the input variables for the CP-odd case with $c_{b} = 1.0, \tilde{c}_{b} = 0.1$ as for signal 
events are presented in Fig.\ref{fig:BDTinputHLLHC2}. The invariant mass of the leading photon and the leading jet system, highest $p_{T}$ b-jet
, and  the difference between the azimuthal angles of $\gamma_{1,2}$ and the $p_{T}$ of the highest $p_{T}$ b-jet  are 
shown as examples of the input variables for the CP-odd BDT. 

\begin{figure}[ht!]
\centering
	\includegraphics[width=0.45\textwidth]{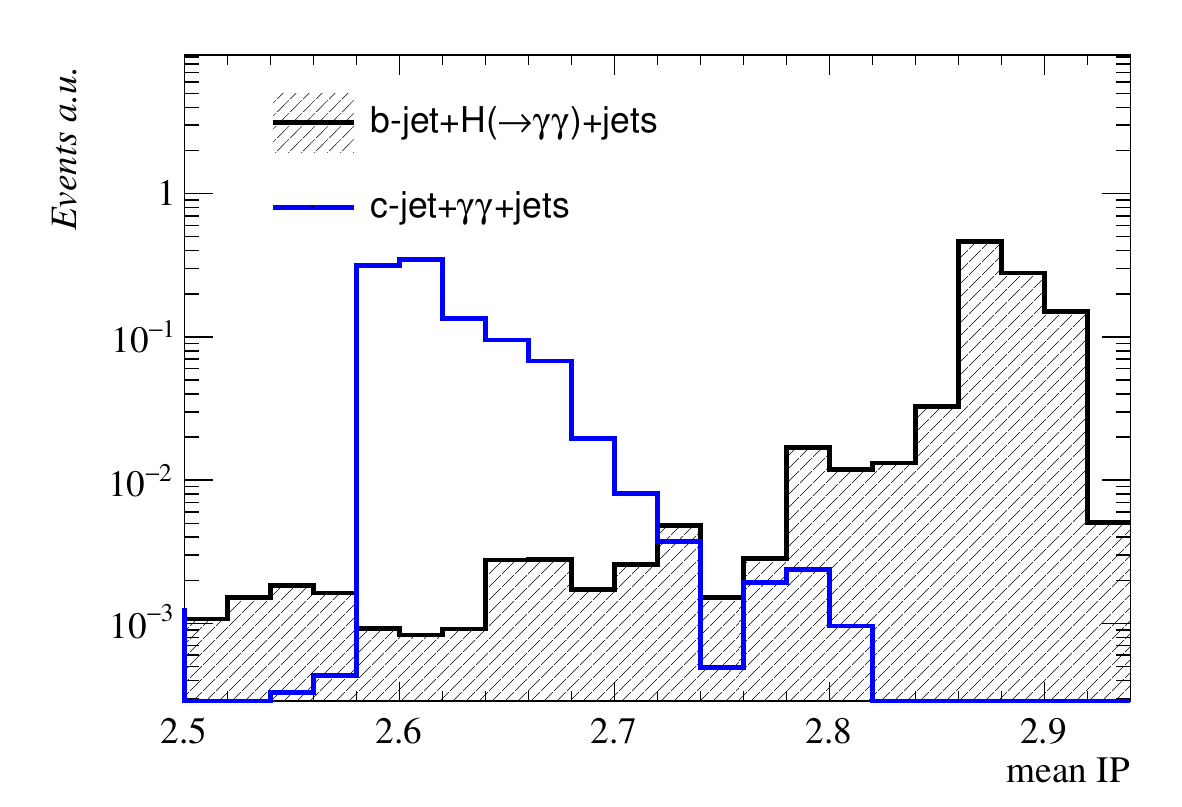}
	\includegraphics[width=0.45\textwidth]{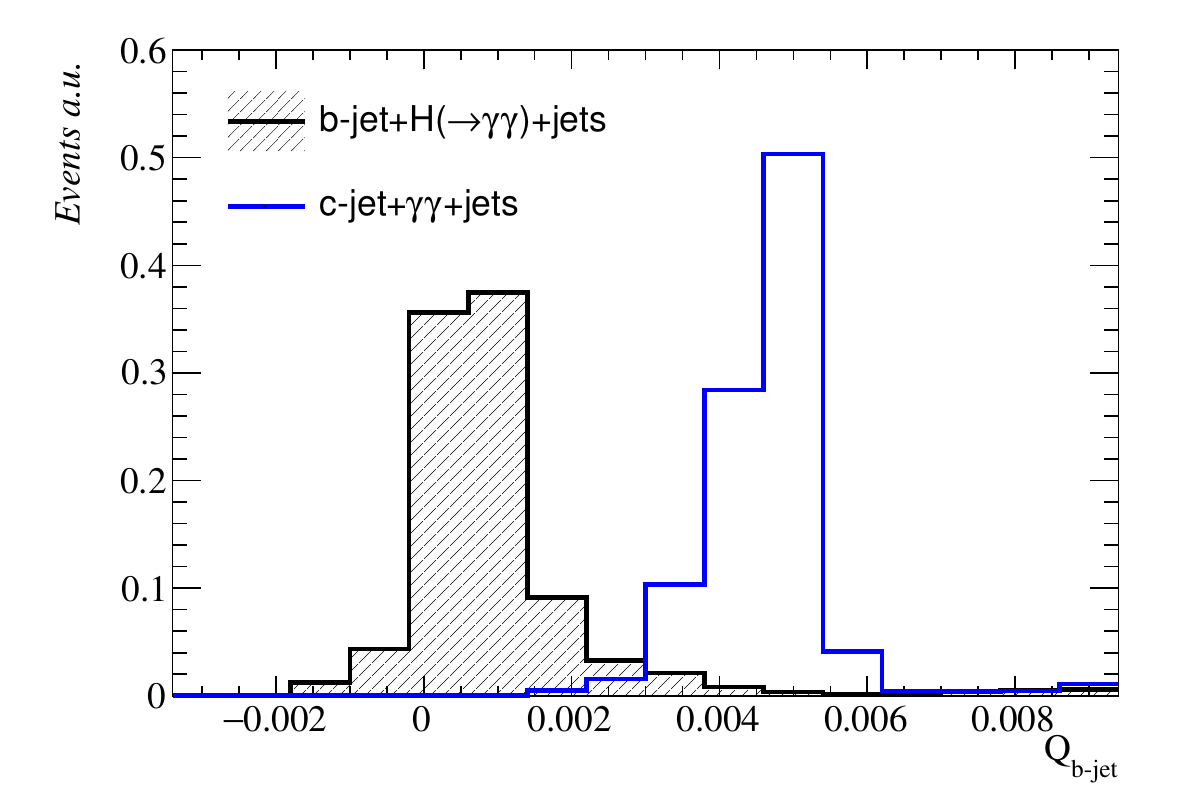}\\
	\caption{Normalised mean IP and $Q_{\rm b-jet}$ distributions for the b-jet+$H(\rightarrow \gamma\gamma)$+jets and the c-jet+$\gamma\gamma$+jets background.
	The definitions of mean IP and $Q_{\rm b-jet}$ are given in Eq.\ref{ippp} and Eq.\ref{qjjj}, respectively.}\label{fig:BDTinputHL-LHC2}
\end{figure}

There is a substantial contribution from the c-jet+$\gamma\gamma$+jets background, primarily arising from the 
elevated misidentification rate of c-quark jets as b-jets. This misattribution is rooted in the reliance of heavy-flavor jet 
identification algorithms on variables linked to the characteristics of heavy-flavor hadrons, such as their lifetimes. 
Notably, heavy-flavor hadrons containing b-quarks exhibit a lifetime on the order of 1.5 ps, whereas c-hadrons have a lifetime of 1 ps or less. 
Consequently, b hadrons typically display displacements ranging from a few millimeters to one centimeter, depending on their 
momentum-values that can align with those of energetic jets containing c-hadrons.
To discern between backgrounds featuring jets originating from c quarks, we utilize the mean impact parameter and the 
charge of the identified b-jet as input variables for the BDT. 
The distribution of these two variables for the b-jet+$H(\rightarrow \gamma\gamma)$+jets and the 
c-jet+$\gamma\gamma$+jets backgrounds is illustrated in Fig.\ref{fig:BDTinputHL-LHC2} for comparative analysis.

\begin{figure}[ht]
	\centering
	\includegraphics[width=0.9\textwidth]{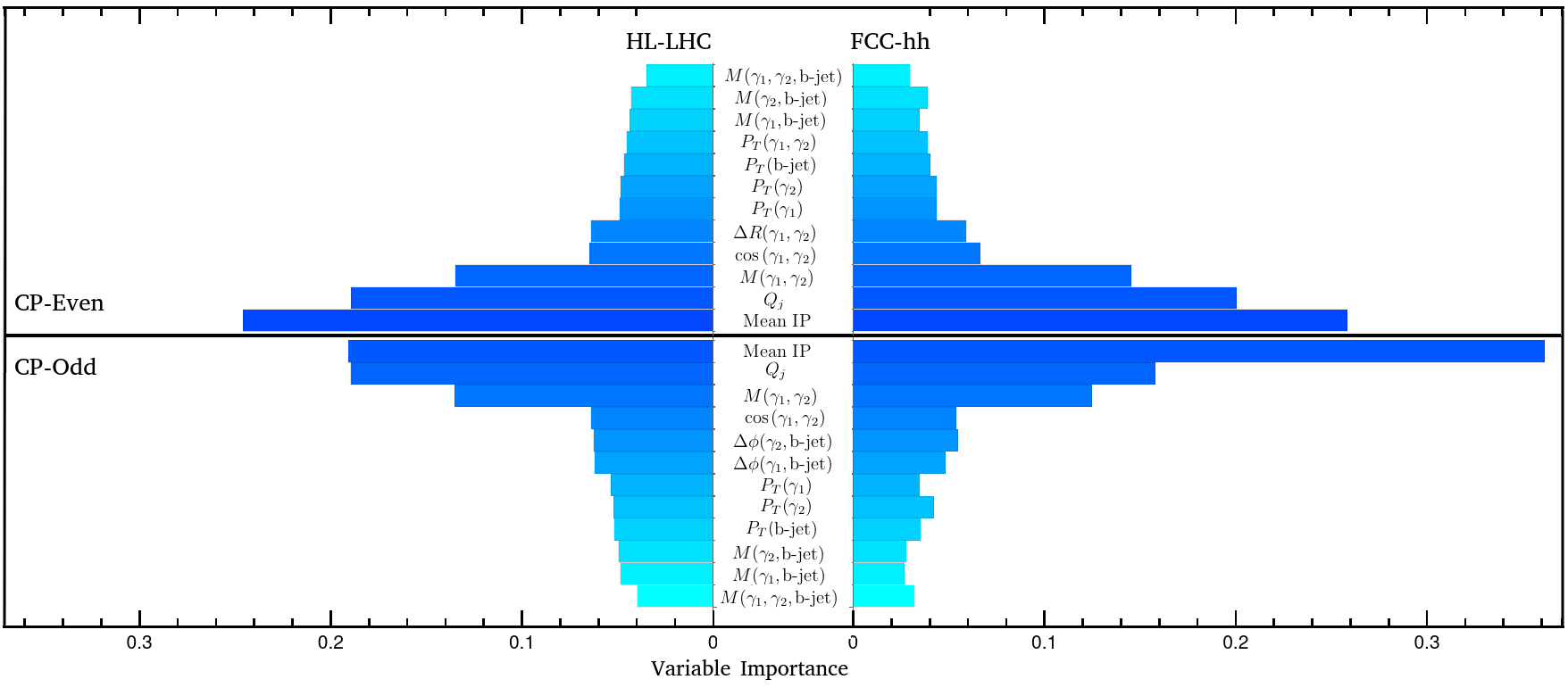}
	\caption{ Input kinematic variables for the Boosted Decision Tree (BDT) and their significance in distinguishing the CP-even signal (top) and CP-odd signal (bottom). 
	On the left, variable importance is illustrated for the LHC, while the right side plot showcases the same analysis for the FCC-hh. }\label{fig:fi}
\end{figure}

It is notable that there is always room for improvement in the analysis, especially by selecting a more optimal set of variables. 
However, the variables utilised in our study have proven to be effective discriminators, as evidenced below.
We employ the relative importance to filter out the most crucial variables while maintaining the accuracy of the BDT. 
By utilising the feature importance, we identify and retain the variables that have the highest significance without compromising the overall accuracy of the BDT model.
The mean IP, b-jet charge, and the invariant mass of the diphoton system, and the cosine between 
the two photons angle emerge as the most significant in distinguishing the signal from the backgrounds
for both the HL-LHC and FCC-hh.
In Figure \ref{fig:fi}, we present the relative importance of each observable based on the 
 separation between the signal and backgrounds.
The ranking is shown for both the HL-LHC and the FCC-hh.
The top panel  illustrates the relevance of kinematic variables as inputs for the BDT in distinguishing CP-even signals, 
 while the bottom panel focuses on CP-odd signals from the main background processes. 
 The left subplot depicts variable importance for the LHC, and the right subplot presents the corresponding analysis for the FCC-hh.

\begin{figure}[h]
\centering
	\includegraphics[width=0.45\textwidth]{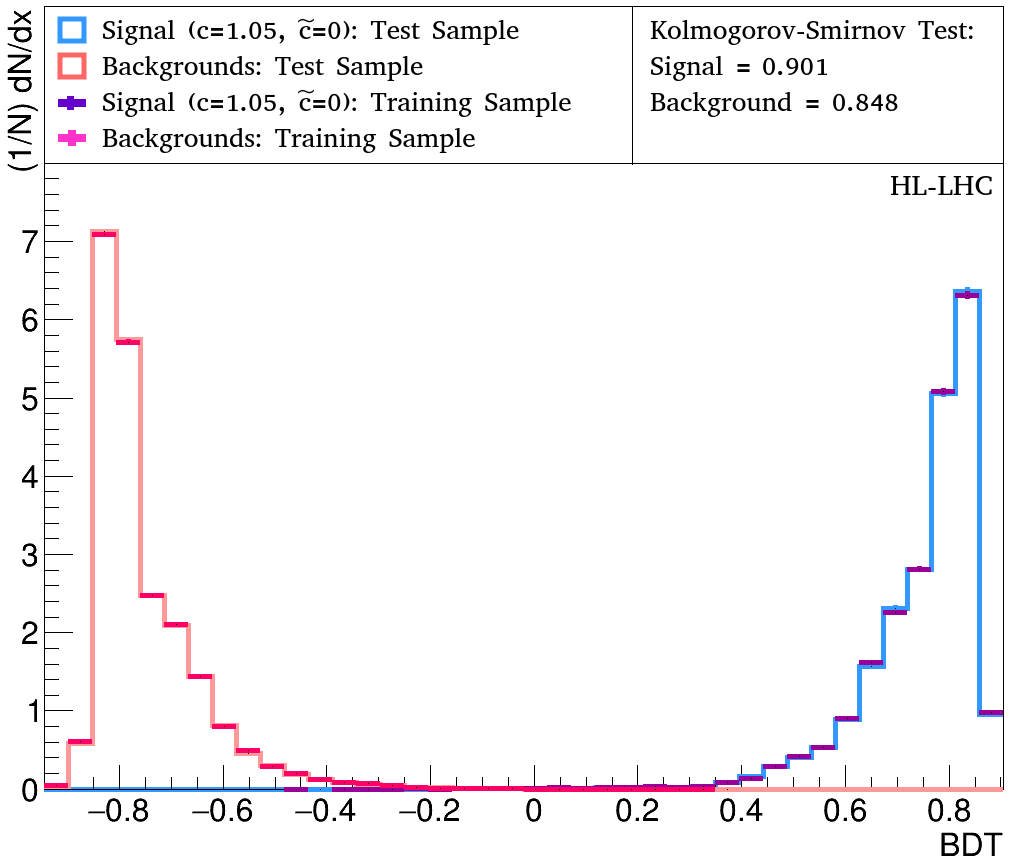}
	\includegraphics[width=0.45\textwidth]{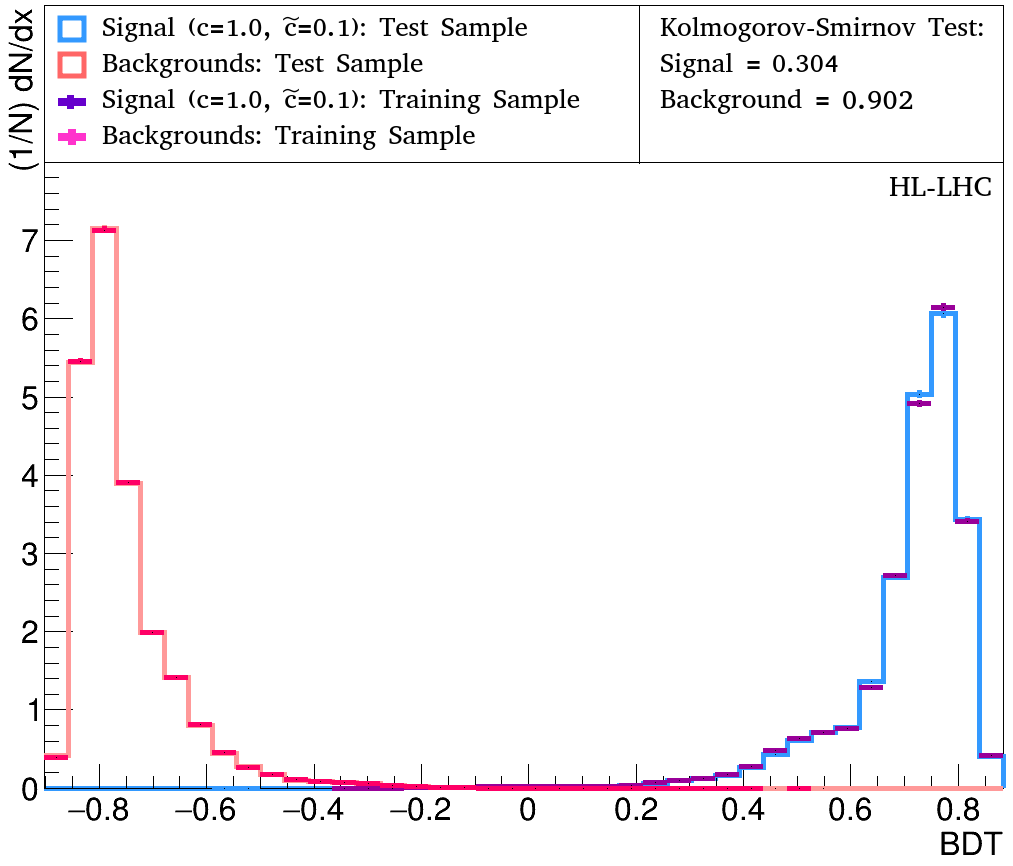}\\
	\includegraphics[width=0.45\textwidth]{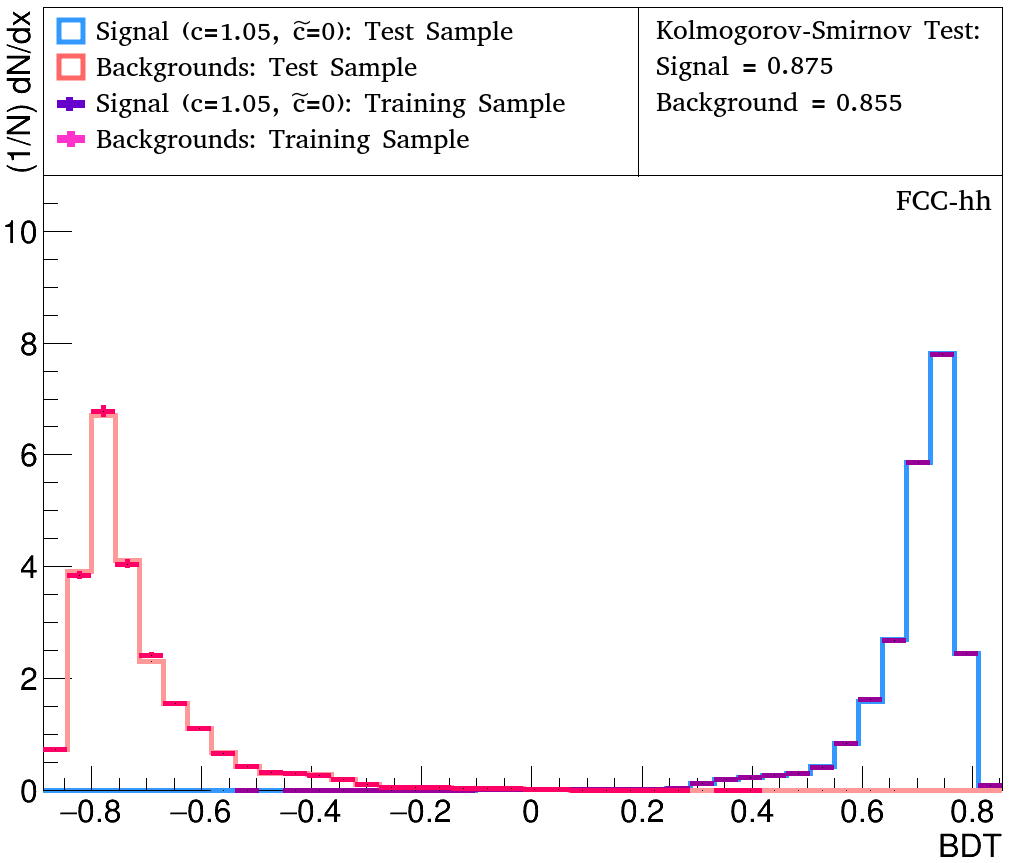}
	\includegraphics[width=0.45\textwidth]{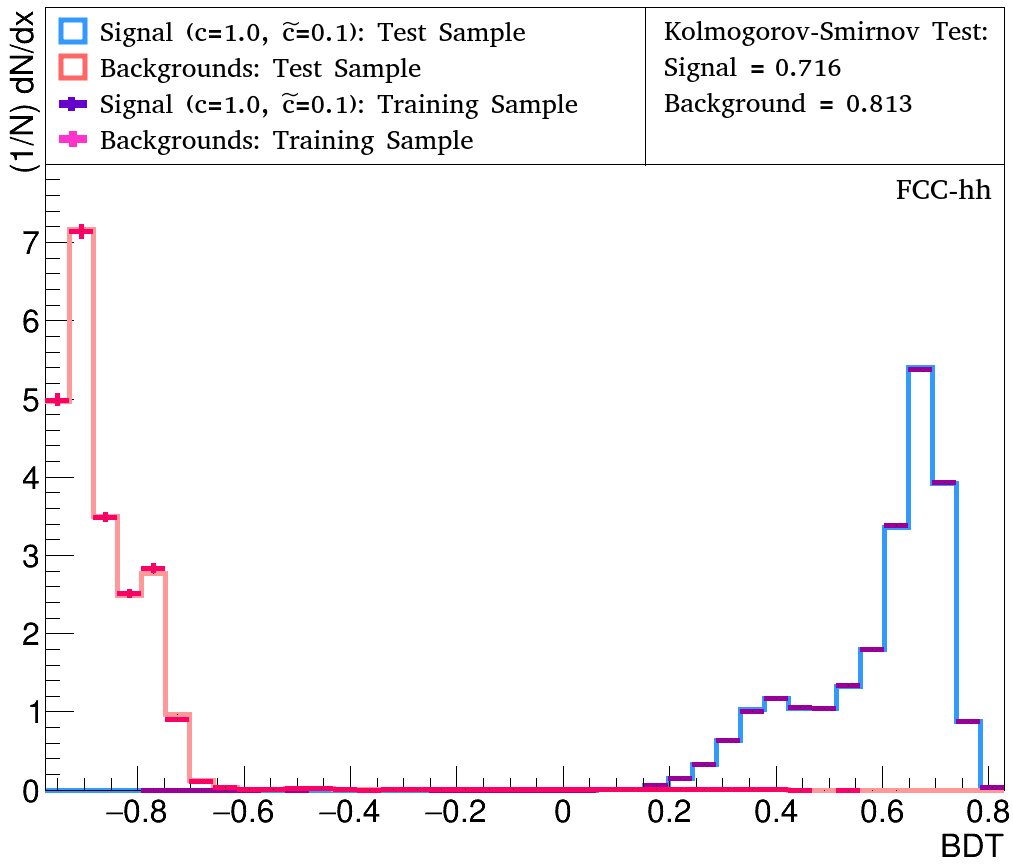}\\
	\caption{The normalized BDT output distributions for signal and background at HL-LHC (top) and FCC-hh (bottom). 
	The distributions are presented for the CP-even scenario with $(c_{b} = 1.05, \tilde{c}_{b} = 0.0)$ (left) and 
	the CP-odd case with $(c_{b} = 1.0, \tilde{c}_{b} = 0.1)$ (right).  The Kolmogorov-Smirnov probabilities are given in the plots as well.}\label{fig:BDT result}
\end{figure}

We proceed with the implementation of the BDT algorithm using the following methodology. 
The datasets containing independent event samples for both the signal and background are 
randomly divided into two equal parts. One part is used to train the BDT algorithm, while the 
other serves as a validation set for both signal and background events.

As mentioned earlier, we employ twelve parameters to train the BDT algorithm separately for CP-even and CP-odd. 
To ensure optimal performance and minimise the risk of overtraining, we take necessary precautions.  
 We have taken explicit measures to prevent overtraining by verifying that the Kolmogorov-Smirnov probability, 
 which measures the similarity between distributions.
Figure \ref{fig:BDT result} displays that BDT output distributions for both HL-LHC (top) and FCC-hh (bottom) and illustrates the Kolmogorov-Smirnov 
probability for both the training and testing samples, demonstrating that neither the signal nor the background 
samples are overtrained. 
To optimise sensitivity, we apply an appropriate cut on the BDT response. 
This cut is determined to maximise the ability to discriminate between the signal and background events. 
By applying this cut, we obtain the corresponding numbers of signal (${\mathbf N_{\rm S}}$) and background (${\mathbf N_{\rm B}}$) events.
Using these event counts, we calculate the sensitivity for parameters $c_b$ and $\tilde{c}_b$.

\section{Results}
\label{sec3}

To determine the statistical significance, denoted as $\mathcal{S}$, we employ the following formula. Given a number of signal events (${\mathbf N_{\rm S}}$)
 and background events (${\mathbf N_{\rm B}}$) at a specific luminosity ($\mathcal{L}$), considering an uncertainty of 
 $\Delta_{\rm B}$ on background, the significance ($\mathcal{S}$) is calculated as \cite{s1,s2}:
 \begin{eqnarray}
 \mathcal{S} = \left[ 2\times \left( (\mathbf N_{\rm S}+\mathbf N_{\rm B})\ln\left[\frac{( \mathbf N_{\rm S}+\mathbf N_{\rm B})( \mathbf N_{\rm B}+\Delta^{2}_{\rm B})}{\mathbf N^{2}_{\rm B}+(\mathbf N_{\rm S}+\mathbf N_{\rm B})\Delta^{2}_{\rm B}} \right]  - \frac{\mathbf N^{2}_{\rm B}}{\Delta^{2}_{\rm B}} \ln\left[1+ \frac{\Delta^{2}_{\rm B} \times \mathbf N_{\rm S} }{\mathbf N_{\rm B}(\mathbf N_{\rm B} +\Delta^{2}_{\rm B} )}  \right] \right) \right]^{1/2}.
 \end{eqnarray}
 
\begin{figure}[h] 
	\centering	
	\includegraphics[width=1.0\textwidth]{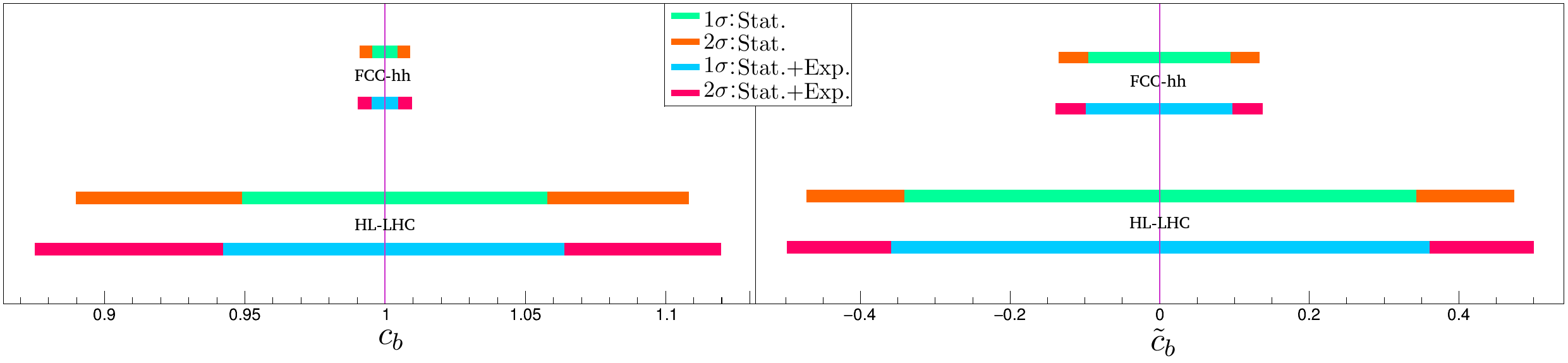}
	\caption{The  $1\sigma $ and $ 2\sigma $ regions for $c_{b}$ and $\tilde{c}_{b}$  by considering only statistical uncertainty and statistical uncertainty plus 
	$25\%$ overall systematic uncertainty at the HL-LHC and FCC-hh.  }\label{fig:lim}
\end{figure}

 In case that  $\Delta_{\rm B} = 0$, $\mathcal{S}$ is reduced to the following formula:
\begin{eqnarray}
\mathcal{S} = \sqrt{2\times \left[ (\mathbf N_{\rm S} + \mathbf N_{\rm B}) \ln(1+\frac{\mathbf N_{\rm S}}{\mathbf N_{\rm B}}) -   \mathbf N_{\rm S} \right]}.            
\end{eqnarray}
In the limit of large number of background events with respect to signal, $\mathcal{S} = \mathbf N_{\rm S}/\sqrt{\mathbf N_{\rm B}}$.
Now, we present the sensitivity reach for both the HL-LHC
and the FCC-hh. In Fig.\ref{fig:lim},  the $1\sigma$ and $2\sigma$ allowed regions
for $c_{b}$ and $\tilde{c}_{b}$ are displayed in Table \ref{tab:lim} and Fig.\ref{fig:lim}.
The integrated luminosity considered for the HL-LHC is 3 ab$^{-1}$, while for the FCC-hh, it is taken as 15 ab$^{-1}$. 
The regions allowed within $1\sigma$ and $2\sigma$ are also displayed, 
accounting for a systematic uncertainty of $25\%$ for background.

\begin{table}[ht!]
	\begin{center}
		\begin{tabular}{|p{0.8cm} |p{0.85cm}|p{1.9cm}|p{5.4cm}|p{5.8cm}|  }  \hline 
			\centering
			$c_{b}/\tilde{c}_{b}$ & $1\sigma/2\sigma$ & Uncertainty &HL-LHC & FCC-hh\\ 
			\hline
			\multirow{4}{*}{\rotatebox[origin=c]{90}{ $ c_{b} \; ( \tilde{c}_{b}= 0) $}} &
			\multirow{2}{*}{1$\sigma $}  &
			Stat. & $ [-1.057 , -0.948] $  $\cup \;[0.949 , 1.057] $ &$ [-1.0045 , -0.995]  $  $\cup \; [0.995 , 1.0045] $\\ \cline{3-5}
			& & Stat.+Exp. & $ [-1.063 , -0.940] $  $ \cup \; [0.941, 1.064]  $ &$ [-1.0048 , -0.996]  $  $\cup \; [0.996 , 1.0048] $\\ \cline{2-5}
			
			& \multirow{2}{*}{2$\sigma $}  & 
			\multirow{1}{*}{Stat.} &
			$ [-1.107 , -0.889] $   $\cup \; [0.889 , 1.108] $ &$ [-1.009 , -0.990]  $   $\cup \;[0.990, 1.009] $\\ \cline{3-5}
			
			& & Stat.+Exp. & $ [-1.120 , -0.873] $  $ \cup \; [0.873 , 1.120]  $ &$ [-1.010 , -0.991]  $  $\cup \; [0.991 , 1.010] $\\ \cline{2-5}
			
			\hline
			\multirow{4}{*}{\rotatebox[origin=c]{90}{$ \tilde{c}_{b} \;  (c_{b} = 1) $}} &
			\multirow{2}{*}{1$\sigma $}  &
			Stat. &
			$[ -0.33 , 0.33] $ &$ [-0.069 , 0.069] $\\ 	\cline{3-5}
			& & Stat.+Exp. &
			$ [-0.35 , 0.35] $ &$ [-0.072 , 0.072] $\\ 	\cline{2-5}
			& \multirow{2}{*}{2$\sigma $} 
			&Stat. & 
			$ [-0.46 , 0.46] $ &$ [-0.097 , 0.097] $\\  \cline{3-5}
			& &Stat.+Exp. &
			$ [-0.49, 0.49] $ &$ [-0.101 , 0.101] $\\  \cline{2-4}
			\hline
		\end{tabular}
	\end{center}
	\caption{The expected $1\sigma$ and $2\sigma$ limits on $c_{b}$ and $\tilde{c}_{b}$ couplings considering only statistical uncertainty and statistical uncertainty plus 
		$25\%$ overall systematic uncertainty at the HL-LHC and FCC-hh with the integrated luminosity of 3 ab$^{-1}$ and 15 ab$^{-1}$.}
	\label{tab:lim}	
\end{table}

\subsection{Bounds on $(c_{b}, \tilde{c}_{b})$ space}

\begin{figure}[h]
	\centering	
	\includegraphics[width=0.8\textwidth]{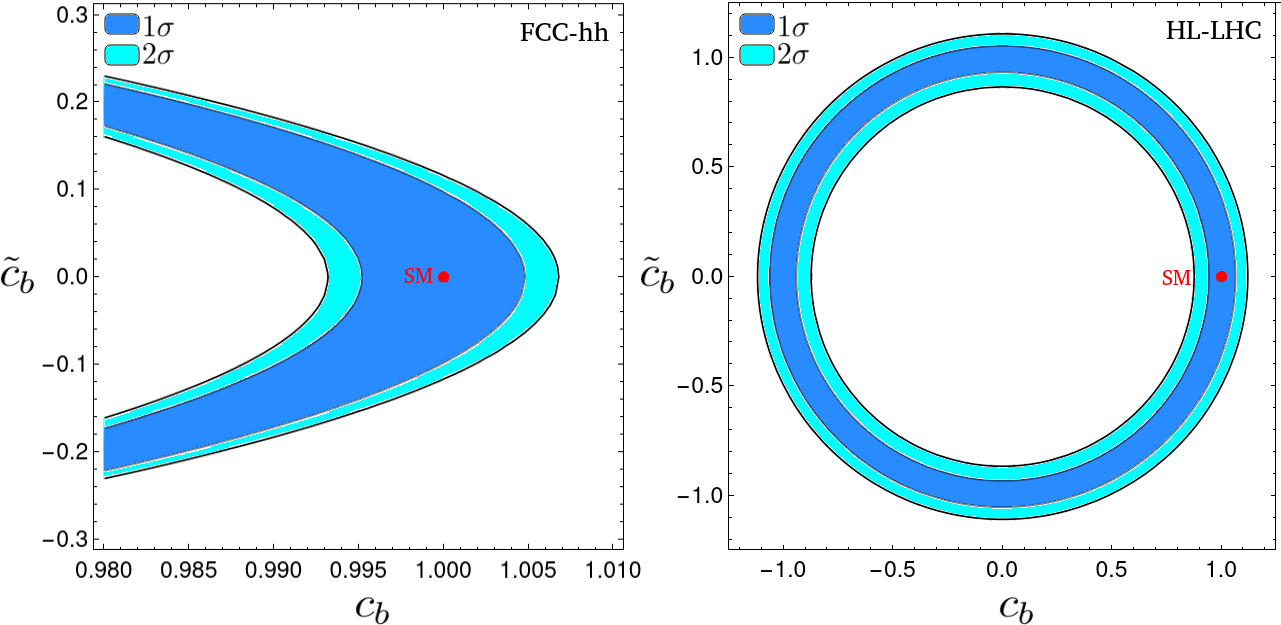}
	\caption{ Expected $2\sigma$ regions in the $c_{b}-\tilde{c}_{b}$ plane obtained for the HL-LHC
        and FCC-hh with the integrated luminosities of 3 ab$^{-1}$ and 15 ab$^{-1}$, respectively. The limits are derived
        assuming an overall uncertainty of $25\%$ on the background estimation. The red circle indicates the SM.}
	\label{contour-result}
\end{figure}

Figure \ref{contour-result} illustrates the anticipated $1\sigma$ exclusion regions within the $c_{b}-\tilde{c}_{b}$ parameter space for both the 
HL-LHC and FCC-hh colliders, taking into account integrated luminosities of 3 ab$^{-1}$ and 15 ab$^{-1}$, respectively. 
These exclusion limits were calculated under the assumption that the kinematics of the signal events remain independent of the
 specific values of the $c_{b}$ and $\tilde{c}_{b}$ couplings. Additionally, the presented limits incorporate an overall systematic
  uncertainty of $25\%$. Upon comparison, it is evident that the limits experience slight enhancements with the increase in 
  center-of-mass energy from the HL-LHC to the FCC-hh. 
  The parameter space region obtained for the FCC-hh exhibits a circular shape, similar to that for the HL-LHC.
  However, due to the compactness of the region delineated for the FCC-hh, we opted to zoom in on a restricted 
  area surrounding the SM value. Consequently, this adjustment resulted in an elliptical appearance, 
  facilitating a more refined visualization of the parameter space vicinity to the SM point.

\section{Exploration of the $Hb\bar{b}$ CP-odd coupling}
\label{seccp}

In this section, we introduce an asymmetry observable that is sensitive to  the magnitude of the 
pseudo-scalar coupling ($\tilde{c}_{b}$). To evaluate this observable, 
we apply the simulation chain and selection criteria described in Section \ref{sec2} for various values of the $\tilde{c}_{b}$ coupling.

\begin{figure}[h] 
	\centering	
	\includegraphics[width=0.55\textwidth]{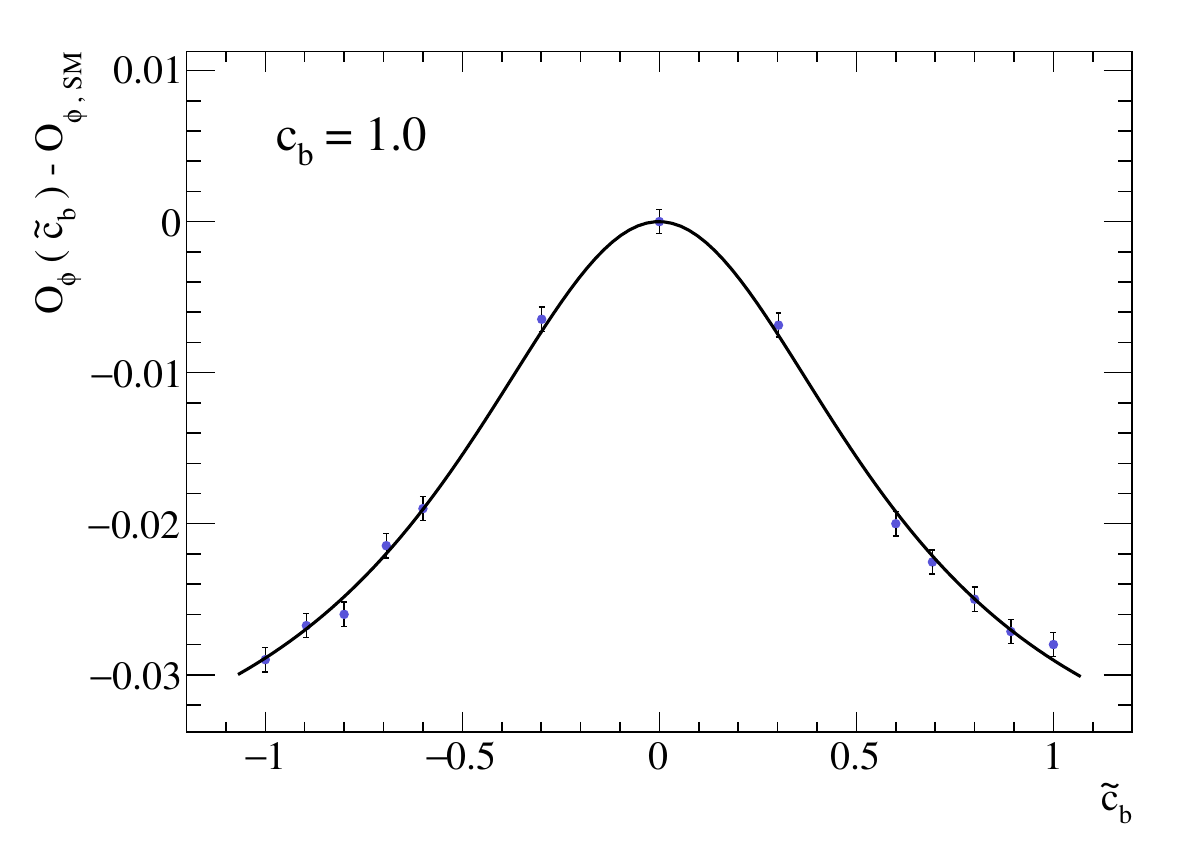}
	\caption{ The variation of the asymmetry from its SM prediction is examined by plotting the difference
	 $\delta\mathcal{O}_{\phi}$ as defined in Eq.\ref{del}. 
	This variation is analyzed as a function of $\tilde{c}_{b}$ at the LHC. The uncertainty is pure statistical. }\label{fig:rr}
\end{figure}

The differential production cross section associated to any CP-mixed 
case of the $Hb\bar{b}$ coupling in H+b+jet signal can be parameterized  according to
the following:
\begin{eqnarray}\label{cscs}
d\sigma = c_{b}^{2}\times d\sigma_{\rm CP-even} +  \tilde{c}_{b}^{2}\times d\sigma_{\rm CP-odd} + c_{b}\times\tilde{c}_{b}\times d\sigma_{\rm int.},
\end{eqnarray}
where $d\sigma_{\rm CP-even}$, $d\sigma_{\rm CP-odd}$, and $d\sigma_{\rm int.}$ are
corresponding to the signal differential cross sections for the CP-even, CP-odd couplings and interference terms, respectively. 
The integration of the interference term in Eq. \ref{cscs} over the whole phase space disappears
because when a CP-even amplitude and a CP-odd amplitude interfere, the resulting interference term oscillates in 
sign across different regions of phase space and the integral of an odd function over a symmetric interval vanishes.
Consequently, the interference term doesn't add anything to the overall rate or to CP-even measurements like 
transverse momenta and invariant masses distributions. 
Instead, it only affects observables designed specifically to measure CP-odd phenomena.
We construct an asymmetry observable from the azimuthal angular distributions of the
final state objects in $pp \rightarrow H (\rightarrow \gamma\gamma)+b+j$ process.  
This observable, $\mathcal{O}_{\phi}$, is sensitive  to
 the CP-violating $\tilde{c}_{b}$ coupling of the $Hb\bar{b}$ interaction.
We define the angular asymmetry  $\mathcal{O}_{\phi}$ with respect to the azimuthal angle as:
\begin{eqnarray}\label{asymm}
\mathcal{O}_{\phi} = \frac{N^{+} - N^{-}}{N^{+} + N^{-}}, 
\end{eqnarray}
where
\begin{eqnarray} \label{rr}
N^{+} = \int_{0}^{\pi} \frac{dN}{d\Delta\phi( H,b)}, ~~ \text{and}~~N^{-} = \int_{-\pi}^{0} \frac{dN}{d\Delta\phi( H,b)},
\end{eqnarray}
where $\Delta\phi( H,b)$ is defined as the azimuthal angle between the Higgs boson
and the highest $p_{\rm T}$ b-quark in the event, where the Higgs boson is reconstructed from the two photons.
 According to the general expression provided in Eq.\ref{cscs}, we anticipate the following functional form for the asymmetry:
\begin{eqnarray}
 \mathcal{O}_{\phi}(c_{b},\tilde{c}_{b}) = \frac{A\times c_{b}^{2}+B\times \tilde{c}_{b}^{2}+ C\times c_{b}\tilde{c}_{b}}{D\times c_{b}^{2}+E\times \tilde{c}_{b}^{2}},
\end{eqnarray}
where $ \mathcal{O}_{\phi}(c_{b}=1.0,\tilde{c}_{b}=0.0)$ is the SM case. We note that the denominator represents the total cross section and thus does not contain an interference term.
Assuming  $c_{b} = 1.0$, the deviation of $\mathcal{O}_{\phi}$ due to the CP-odd coupling from the SM has the following form:
\begin{eqnarray} \label{del}
\delta\mathcal{O}_{\phi}(\tilde{c}_{b}) = \mathcal{O}_{\phi}(c_{b}=1.0,\tilde{c}_{b}) - \mathcal{O}_{\phi}(c_{b}=1.0,\tilde{c}_{b}=0.0) = \frac{B' \times \tilde{c}_{b}^{2}+ C' \times \tilde{c}_{b}}{1.0+E' \times \tilde{c}_{b}^{2}},
\end{eqnarray}
where parameters $B' \equiv (BD-AE)/D^{2}$,  $C' \equiv C/D$,  and $E' \equiv E/D$. 
To illustrate the sensitivity of the asymmetry, we explore the relationship between the $\delta\mathcal{O}_{\phi}$ and 
 $\tilde{c}_{b}$ coupling. Several Monte-Carlo simulated samples consisting of 500K events is  analyzed to discern the 
degree to which the asymmetry is influenced by the presence of the $\tilde{c}_{b}$ coupling.
The difference between the asymmetry and its SM value, $\delta\mathcal{O}_{\phi}$, 
is plotted against $\tilde{c}_{b}$ at the LHC, 
as illustrated in Figure \ref{fig:rr}. In the  plot, the value of $c_{b}$ is fixed at the SM value of $1.0$ and the uncertainty depicted is purely statistical.
By conducting a fit, we derive the following result: $B' = -0.09\pm 0.003, C' = 0.0006\pm 0.0005$, and $E' = 2.9\pm 0.3$.
It is evident from the plot that the $\mathcal{O}_{\phi}$ exhibits sensitivity to  the magnitude 
of CP-odd  coupling. As $|\tilde{c}_{b}|$ increases, $\delta\mathcal{O}_{\phi}$ falls below the SM value.
It is noteworthy that $\delta\mathcal{O}_{\phi}$ receives a minor impact from the interference 
term because of  the very small coefficient of linear term of $\tilde{c}_{b}$ with respect to the term $\tilde{c}_{b}^{2}$, i.e. $|C'| \ll |B'|$.
As a result, the bounds which will be derived on $\tilde{c}_{b}$ using $\mathcal{O}_{\phi}$ are 
expected to exhibit only a minimal degree of asymmetry.

We also investigate the sensitivity of the asymmetry as a probe of CP-violating 
couplings at the HL-LHC and FCC-hh. To quantify this, all selection criteria presented in section \ref{evsel} are followed.
 We assess the statistical significance
 in the measurement of the asymmetry $\mathcal{O}_{\phi}$ as  
  $S_{\mathcal{O}_{\phi}} = \mathcal{O}_{\phi}/\Delta \mathcal{O}_{\phi}$, where $\Delta \mathcal{O}_{\phi}$ is:
\begin{eqnarray}
\Delta \mathcal{O}_{\phi} = \sqrt{\frac{1-\mathcal{O}^{2}_{\phi}}{\sigma_{\rm SM}\times \mathcal{L}}},
\end{eqnarray}
where $\sigma_{\rm SM}$ is the SM cross section and $\mathcal{L}$ is the integrated luminosity.
The signal significance $S_{\mathcal{O}_{\phi}}$ is dependent on $c_{b}$ and $\tilde{c}_{b}$ and has the following form:
\begin{eqnarray}
S_{\mathcal{O}_{\phi}}(c_{b},\tilde{c}_{b}) = \frac{\mathcal{O}_{\phi} - \mathcal{O}_{\phi, \rm SM}}{\sqrt{1-\mathcal{O}^{2}_{\phi, \rm SM}}}\times \sqrt{\sigma_{\rm SM}\times \mathcal{L}}.
\end{eqnarray}

\begin{figure}[h] 
	\centering	
	\includegraphics[width=0.45\textwidth]{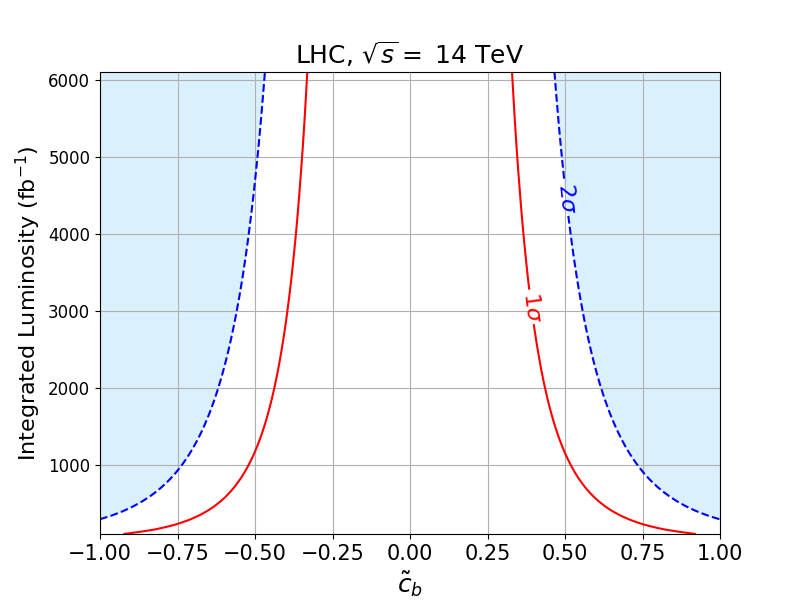}
		\includegraphics[width=0.45\textwidth]{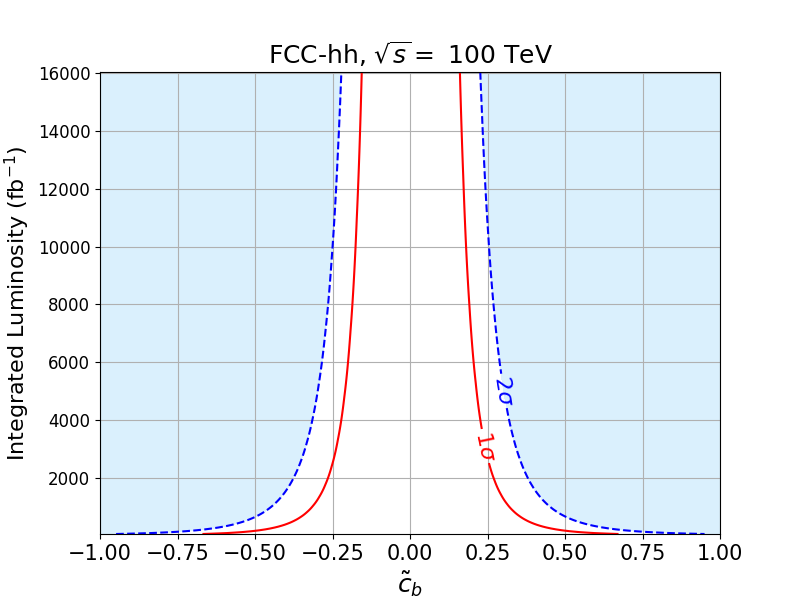}
	\caption{The  $1\sigma $ and $ 2\sigma $ regions versus the integrated luminosity for $\tilde{c}_{b}$ for the HL-LHC (left) and FCC-hh (right).  
	The regions are obtained considering only statistical uncertainty.  }\label{fig:cnt}
\end{figure}

As our focus lies on the CP-odd coupling, $c_{b}$ is set to its SM value and we concentrate on $\tilde{c}_{b}$.
In Figure \ref{fig:cnt}, the $1\sigma $ and $ 2\sigma $ regions of $\tilde{c}_{b}$ are depicted versus the integrated luminosity 
for both the HL-LHC and FCC-hh. 

As observed, the $1\sigma$ ($2\sigma$) region of $\tilde{c}_{b}$
 is accessible down to $ -0.40 \leq \tilde{c}_{b} \leq 0.38 ~(-0.62 \leq \tilde{c}_{b} \leq 0.61)$ with an integrated luminosity of 
 3000 fb$^{-1}$ at the HL-LHC and down to $(-0.16 \leq \tilde{c}_{b} \leq 0.15~(-0.23 \leq \tilde{c}_{b} \leq 0.22))$ with an integrated luminosity of 15000 fb$^{-1}$ at the FCC-hh.

\section{Summary and conclusions}
\label{sec4}

This study focused on investigating the sensitivity of the $H+b+\text{jets}$ process at the HL-LHC and FCC-hh colliders in order to 
explore the CP-even and CP-odd couplings of $Hb\bar{b}$. The analysis involved constraining the new physics couplings, $c_b$ 
and $\tilde{c}_b$, by conducting a search through the $H+b+\text{jets}$ channel and studying the subsequent decay of the Higgs boson
 into two photons. Monte Carlo event generation was employed to simulate the signal and relevant background processes, 
 and detector effects were taken into account. To distinguish the signal from the background, a carefully selected set of 
 discriminating variables was analysed using a multivariate technique, in particular, by employing Boosted Decision Trees (BDTs).
 The expected $1\sigma$ and $2\sigma$ limits on $c_b$ and $\tilde{c}_b$ 
were obtained, and the exclusion region in the $c_b$-$\tilde{c}_b$ plane were determined. 
These limits correspond to integrated luminosities of 3 ab$^{-1}$ and 15 ab$^{-1}$ for the HL-LHC and FCC-hh, respectively.

Comparing the obtained limits with the current experimental bounds, it was found that a significant portion of the 
unexplored $c_b$-$\tilde{c}_b$ parameter space could be accessed through this analysis. 
By assuming one non-zero coupling at a time for comparison, the $c_b$ coupling was found to range from $[-1.2,-1.17] \cup [0.88,1.12]$ 
based on the LHC Higgs signal strength data at $90\%$ confidence level (CL). The direct predicted ranges from this study 
were $[-1.12,-0.873] \cup [0.873,1.12]$ for the HL-LHC and $[-1.010,-0.991] \cup [0.991,1.010]$ for the FCC-hh.

Regarding the $\tilde{c}_b$ parameter, the current limits were within the range $[-0.5,0.5]$ based on the LHC Higgs signal 
strength data and $[-0.26,0.26]$ from electron EDM at $90\%$ CL. The direct measurement from this analysis excluded the 
range $[-0.49,0.49]$ for the HL-LHC and $[-0.10,0.10]$ for the FCC-hh at the $2\sigma$ level.
Both the HL-LHC and FCC-hh demonstrated the potential to provide limits on $c_b$ and $\tilde{c}_b$ at the same order 
of magnitude as or even better than those obtained from indirect bounds. 

We also introduced a novel asymmetry tailored to investigate CP violation 
within the $Hb\bar{b}$ coupling, applying exclusively lab-frame momenta. 
Our analysis underscored the effectiveness of this asymmetry in constraining
the CP-odd couplings of the bottom quark Yukawa couplings.

Upon contrasting the findings of this study, 
it becomes apparent that the $H+b+\text{jets}$ process emerges as a powerful and direct avenue for investigating both 
the CP-even and CP-odd aspects of the $Hb\bar{b}$ interactions at proton-proton colliders. 
This efficacy is primarily attributed to the incorporation of a more extensive final state.
There are potential avenues to improve the results. 
Firstly, incorporating complete next-to-leading order predictions for the $H+b+\text{jet}$ process, including loop level diagrams, 
can yield more accurate and reliable outcomes.
Secondly, to enhance sensitivity and statistical significance, 
the inclusion of additional decay modes of the Higgs boson, such as $WW$ and $ZZ$
needs to be considered.

\vspace{0.5 cm}
{\bf Acknowledgement}\\

The authors would like to thank M. Ebrahimi for her fruitful discussions in multivariate analysis algorithms
and M. Zaro for his valuable guidances in generating events with MadGraph. 
%
%

\end{document}